\documentclass[amsmath,amssymb,twocolumn,aps,pra,10pt,floatfix,superscriptaddress,a4paper]{revtex4-1}

\usepackage{url}
\usepackage{graphicx} 
\usepackage{multirow}
\usepackage{float} 
\usepackage{wrapfig} 
\usepackage{lipsum} 
\usepackage{booktabs}
\usepackage[normalem]{ulem}

\usepackage[utf8]{inputenc}
\usepackage{amsmath}
\usepackage{amsfonts}
\usepackage{amssymb}
\usepackage{fancyhdr}
\usepackage{amsthm}

\usepackage{enumitem}
\usepackage{amsmath}
\usepackage{braket}
\usepackage{graphicx}
\usepackage{pgfplots}
\usepackage{array}
\usepackage{placeins}

\usepackage{mathtools}

\graphicspath{ {pics/} }

\usepackage[vcentermath]{youngtab}

\usepackage[colorlinks=true,linkcolor={blue}]{hyperref}
\usepackage[colorinlistoftodos]{todonotes}

\newcommand{\mr}[1]{\mathrm{#1}}
\def\Id{1\!\mathrm{l}}

\DeclareMathOperator{\Tr}{Tr}

\newcommand{\crop}[2]{{\hat{#1}}_{#2}^{\dagger}}

\theoremstyle{definition}

\theoremstyle{plain}

\theoremstyle{remark}

\newcommand{\iu}{{i\mkern1mu}}

\newcommand{\ytwo}{\scalebox{.3}{\yng(2)}}
\newcommand{\yoneone}{\scalebox{.3}{\yng(1,1)}}
\newcommand{\yoneoneone}{\scalebox{.3}{\yng(1,1,1)}}
\newcommand{\ytwoone}{\scalebox{.3}{\yng(2,1)}}
\newcommand{\ythree}{\scalebox{.3}{\yng(3)}}

\DeclarePairedDelimiter\abs{\lvert}{\rvert}%
\DeclarePairedDelimiter\norm{\lVert}{\rVert}%

\makeatletter
\let\oldabs\abs
\def\abs{\@ifstar{\oldabs}{\oldabs*}}
\let\oldnorm\norm
\def\norm{\@ifstar{\oldnorm}{\oldnorm*}}
\makeatother

\begin{document}
\title{Discriminating distinguishability}
\author{Stasja \surname{Stanisic}}
\affiliation{Quantum Engineering Technology Labs, H. H. Wills Physics Laboratory and Department of Electrical \& Electronic Engineering, University of Bristol, UK}
\affiliation{Quantum Engineering Centre for Doctoral Training, University of Bristol, UK}
\author{Peter S. \surname{Turner}}
\affiliation{Quantum Engineering Technology Labs, H. H. Wills Physics Laboratory and Department of Electrical \& Electronic Engineering, University of Bristol, UK}

\date{\today}

\begin{abstract}
Particle distinguishability is a significant challenge for quantum technologies, in particular photonics where the Hong-Ou-Mandel (HOM) effect clearly demonstrates it is detrimental to quantum interference.
We take a representation theoretic approach in first quantisation, separating particles' Hilbert spaces into degrees of freedom that we control and those we do not, yielding a quantum information inspired bipartite model where distinguishability can arise as correlation with an environment carried by the particles themselves.
This makes clear that the HOM experiment is an instance of a (mixed) state discrimination protocol, which can be generalised to interferometers that discriminate unambiguously between ideal indistinguishable states and interesting distinguishable states, leading to bounds on the success probability of an arbitrary HOM generalisation for multiple particles and modes. 
After setting out the first quantised formalism in detail, we consider several scenarios and provide a combination of analytical and numerical results for up to nine photons in nine modes.
Although the Quantum Fourier Transform features prominently, we see that it is suboptimal for discriminating completely distinguishable states.

\end{abstract}


\maketitle

\section{Introduction}
\label{sec:introduction}

Interference lies at the heart of quantum mechanics, and thus its promise of fundamental advantages over non-quantum technologies, with far-reaching ramifications in communication, metrology, simulation and computation.  The nemesis of quantum interference is distinguishability, with the Hong-Ou-Mandel (HOM) effect~\cite{Hong1987} being a prototypical example.
Recent advances in scaling linear optics for universal quantum computation~\cite{Knill2001,Carolan2015,Rudolph2016}, and the race to demonstrate quantum computational `supremacy' via analog computations that sample the scattering amplitudes of multipartite states~\cite{Aaronson2011,Spring2013,Broome2013,Crespi2013,Tillmann2013, Bentivegna2015,Wang2017,Neville2017,Viggianiello2017}, highlight the need for a thorough understanding of distinguishability in multimode quantum interference~\cite{Tichy2014,Tichy2015,Menssen2017,Shchesnovich2015,Shchesnovich2017,Tillmann2014,Laibacher2015,Rohde2015,Renema2017,Brunner2018}. 

Rather than the usual second quantized approach, we can gain insight by bringing quantum information concepts to bear in first quantization~\cite{Popescu2007,Adamson2008,Oszmaniec2018,Moylett2018}. 
Distinguishability can then be modelled, for example, as entanglement between controlled and uncontrolled degrees of freedom of individual particles, with loss of interference being caused by the decoherence that results when the uncontrolled Hilbert space is marginalized.
This can be formalized by observing that bosonic (and fermionic) Fock states of two (sets of) degrees of freedom can have natural Schmidt decompositions, corresponding to so called unitary-unitary duality in many-body physics~\cite{Rowe2012}.

An example of a pertinent idea from quantum information is state discrimination~\cite{Chefles1998,Bergou2004,Barnett2009}; we start by showing how this reproduces the well known HOM distinguishability test for two particles.
In principle the formalism accommodates any number of particles and modes, and we show how this generalises for multimode quantum interference, taking a representation theoretic approach (Sections~\ref{sec:background} and~\ref{sec:manyParticles}); this complements a number of generalizations in the literature~\cite{Campos2000,Walborn2003,Lim2005,Englert2008,Tichy2010,Garcia2013,Crespi2015,Rigovacca2016,Brod2018}.
We set up the state discrimination problem in the linear optical framework, assuming we have access to passive transformations (networks of phaseshifters and beamsplitters) and projective measurements via photon number counting detectors (Sec.~\ref{sec:discrimination}).
This restriction on the allowed measurements yields a highly nontrivial constraint on the mixed state discrimination scenario -- this new problem is what we study here.
In particular, the optimisation problem that results is nonlinear, as is usually the case in multiphoton interferometry~\cite{Vanmeter2007}, necessitating numerical techniques described in Sec.~\ref{subsec:numerical}.

The results are as follows: in Sec.~\ref{subsec:bounds} we present two general upper bounds valid for any photon number $N$ when discriminating (i) a state with a single distinguishable photon from the completely indistinguishable state, and (ii) the completely distinguishable from the completely indistinguishable state; in Sec.~\ref{subsec:two} we show why the HOM test is the only test of distinguishability for arbitrary states of two photons, and demonstrate the generality of the formalism by considering three photons in two modes; in Sec.~\ref{subsec:three} we use a mix of analytical and numerical techniques to argue the optimality of a balanced three mode network (tritter) as a discriminator for both completely distinguishable and singly distinguishable states;  in Sec.~\ref{subsec:moreOBE} we look at discrimination of singly distinguishable states with higher photon numbers up to $N=9$ and show that the quantum Fourier transform (QFT) saturates the bound, suggesting it is the optimal interferometer for all $N$; finally in Sec.~\ref{subsec:fourAndFive} we look at the discrimination of completely distinguishable states with higher photon numbers and give examples of the best known interferometers up to $N=8$, found by observing a pattern emerging from the optimisations. 
Most of these results are summarised in Table~\ref{tab:CDopt}.
Although not surprising that the QFT features heavily, the results show that it is not optimal for discriminating completely distinguishable states, motivating the search for optimal discriminating networks for other states of interest.

\section{Motivation}
\label{sec:background}

\subsection{Hong-Ou-Mandel interference}
\label{subsec:HOM}
We will use the HOM scenario as an example that sets out the main features of our distinguishability model, and its relationship to state discrimination.
Each HOM photon has two pertinent degrees of freedom: one is spatial, namely the interferometer arms, and the other is temporal, namely the time of arrival.
We are usually interested in the case where it is the spatial degree of freedom over which we have control (via interferometry), and so we call this the `System' degree of freedom.
We interpret the temporal degree of freedom as a `Label' -- in general this would include all the particles' degrees of freedom which we do not control.
Since complete control of the System includes the possibility of putting photons in the same spatial mode, we view the Label as determining the particles' distinguishability, via correlations between the System and Label degrees of freedom.
In a real HOM experiment we are interested in preparing situations with varying distinguishability, so we do in fact manipulate the temporal Label degree of freedom as well, but for applications we usually think of the System-Label correlations as having been determined by means beyond our control.

The HOM scenario has two spatial System modes which we will call ``top'' and ``bottom'' ($s=\uparrow,\downarrow$), and two photons, requiring two temporal Label modes that we will call ``early'' and ``late'' ($l=\leftarrow,\rightarrow$).
(Note that these symbols will need to be ordered -- we have avoided the obvious choice of $s$ and $l=1,2$ to reduce confusion with other indices in this section; in Sec.~\ref{subsec:SchurWeyl} we will revert to integers for the general case.)
Photon creators are written as $\hat{a}^\dag_{s l}$~\footnote{For example $\hat{a}^\dag_{s l} = \int \mathrm{d}\omega \, f_l(\omega) \hat{a}^\dag_{s}(\omega)$ where $f_l$ is a spectral envelope function indexed by $l$.}, giving rise to Fock states which we can write as arrays where rows correspond to System modes and columns to Label modes.
An example of a completely distinguishable two photon state is
\begin{align}
\ket{\psi_\mr{d}} &= \hat{a}^\dag_{\uparrow\leftarrow} \hat{a}^\dag_{\downarrow\rightarrow} \ket{\mr{vac}} = \Ket{\begin{matrix} 1&0\\0&1 \end{matrix}} , \label{eq:d}
\end{align}
with an early photon in the top arm and a late one in the bottom, while
\begin{align}
\ket{\psi_\mr{i}} &= \hat{a}^\dag_{\uparrow\leftarrow} \hat{a}^\dag_{\downarrow\leftarrow} \ket{\mr{vac}} = \Ket{\begin{matrix} 1&0\\1&0 \end{matrix}} \label{eq:i}
\end{align}
corresponds to an indistinguishable state where both photons are early.

Ideally an interferometer acts only upon the System, corresponding to a unitary transformation on the two spatial modes
\begin{align}
\hat{a}^\dag_{s l} \mapsto \sum_{t} \hat{a}^\dag_{t l} U_{ts} . \label{eq:fund}
\end{align}
Here $U$ is a $2 \times 2$ unitary matrix corresponding to the two port interferometer, sometimes called the transfer matrix.
We assume that the interferometer acts trivially upon the Label modes (the photons remain early or late), corresponding to the $2 \times 2$ identity transfer matrix $\Id$.
For a suitable choice of ordering of the four possible creators, the full $4 \times 4$ transfer matrix acting on all four modes ($\uparrow\leftarrow,\uparrow\rightarrow,\downarrow\leftarrow,\downarrow\rightarrow$) is given by
\begin{align}
U \otimes \Id. \label{eq:bipart}
\end{align}
It is tempting to interpret the tensor product in Eq.~(\ref{eq:bipart}) as that between the System and the Label.
A quantum information theoretic approach to distinguishability would then ignore (trace out) the Label, arriving at reduced states on the System where all the nontrivial transformations and measurements occur.
However, this is not the tensor product structure of the four harmonic oscillators in the second quantized model, and so one cannot marginalise, for example, the columns in Eqs.~(\ref{eq:d},\ref{eq:i}).
In order to trace out the Label we will use a first quantized description.

Second quantized Fock states can be related to first quantized single particle states as follows.
Viewing each excitation of our four mode aggregate as a particle with four available states ($\uparrow\leftarrow,\uparrow\rightarrow,\downarrow\leftarrow,\downarrow\rightarrow$), and recognizing that as bosons the total state must be symmetric under particle exchange, we have a one-to-one relationship between the Fock states of two bosons in four modes and symmetric states of two four-dimensional particles, (qu$d$its, here with $d=4$).
Applying this procedure to the indistinguishable state of Eq.~(\ref{eq:i}), we have
\begin{align}
\ket{\psi_\mr{i}}
&=\Ket{\begin{matrix} 1&0\\1&0 \end{matrix}} \label{eq:2ndind}\\
&=\mr{Sym} \left( \ket{\uparrow\leftarrow}_1 \ket{\downarrow\leftarrow}_2 \right) \label{eq:Symi}\\ 
&=\frac{1}{\sqrt{2}}\left( \ket{\uparrow\leftarrow}_1 \ket{\downarrow\leftarrow}_2 + \ket{\downarrow\leftarrow}_1 \ket{\uparrow\leftarrow}_2 \right) \\
&=\frac{1}{\sqrt{2}}\left( \ket{\uparrow\downarrow}_\mr{S}+\ket{\downarrow\uparrow}_\mr{S} \right) \ket{\leftarrow\leftarrow}_\mr{L} , \label{eq:1stind}
\end{align}
where the subscripts 1 and 2 have been used as (fictitious) particle labels that get permuted, and we have rearranged the tensor product structure in the last line to arrive at a state in the S(ystem)$\otimes$L(abel) basis.
Similarly, one finds
\begin{align}
\ket{\psi_\mr{d}} 
&= \Ket{\begin{matrix} 1&0\\0&1 \end{matrix}} \\
&=\mr{Sym} \left( \ket{\uparrow\leftarrow}_1 \ket{\downarrow\rightarrow}_2 \right) \label{eq:Symd}\\ 
&= \frac{1}{\sqrt{2}} \ket{\uparrow\downarrow}_\mr{S} \ket{\leftarrow\rightarrow}_\mr{L} + \frac{1}{\sqrt{2}} \ket{\downarrow\uparrow}_\mr{S} \ket{\rightarrow\leftarrow}_\mr{L} . \label{eq:1stdis}
\end{align}
We see that Eq.~(\ref{eq:1stind}) is in a product state (Schmidt rank 1) of System and Label~\footnote{The System state itself is entangled with respect to the impractical particle tensor product -- this has been referred to as ``free'' entanglement by Aaronson.}, so the Label states are uncorrelated to the System states; learning the Label does not allow one to learn anything about the System, as expected for indistinguishable particles.
Equation~(\ref{eq:1stdis}) is entangled (Schmidt rank 2), with the System states perfectly correlated to the Labels ($\uparrow$ to $\leftarrow$ and $\downarrow$ to $\rightarrow$), making the photons completely distinguishable.

It will be useful to rewrite states of both the System and Label according to their permutation symmetry.
Schur-Weyl duality~\cite{Rowe2012, Harrow2005} ensures that this basis also has good quantum numbers for the unitary group action of the interferometer, in this case U$(2)$~\footnote{We need not be concerned with the inconsequential difference between unitary groups U$(d)$ and special unitary groups SU$(d)$.}.
The irreducible representations (irreps) of U(2) are well known, and for only two particles Young diagrams provide a compact notation for the basis states that carry these irreps; they are (for arbitrary, ordered single particle quantum numbers $x,y$) the symmetric triplet 
\begin{align}
\Ket{\scriptsize\young(xx)\,} &= \ket{xx} , \label{eq:11}\\
\sqrt{2}\Ket{\scriptsize\young(xy)\,} &= \ket{xy} + \ket{yx} , \\
\Ket{\scriptsize\young(yy)\,} &= \ket{yy} ,
\end{align}
and the antisymmetric singlet 
\begin{align} 
\sqrt{2}\Ket{\scriptsize\young(x,y)} &= \ket{xy} - \ket{yx} . \label{eq:00}
\end{align}
We can now rewrite Eqs.~(\ref{eq:1stind}, \ref{eq:1stdis}) as
\begin{align}
\ket{\psi_\mr{i}} = \Ket{\begin{matrix} 1&0\\1&0 \end{matrix}} &= \Ket{\scriptsize\young(\uparrow \downarrow)\,}_\mr{S} \Ket{\scriptsize\young(\leftarrow \leftarrow)\,}_\mr{L} \label{eq:ind} , \\ 
\ket{\psi_\mr{d}} = \Ket{\begin{matrix} 1&0\\0&1 \end{matrix}} &= \frac{1}{\sqrt{2}} \Ket{\scriptsize\young(\uparrow \downarrow)\,}_\mr{S} \Ket{\scriptsize\young(\leftarrow \rightarrow)\,}_\mr{L} 
+ \frac{1}{\sqrt{2}}  \Ket{\scriptsize\young(\uparrow,\downarrow)\,}_\mr{S}  \Ket{\scriptsize\young(\leftarrow,\rightarrow)\,}_\mr{L}.\label{eq:dis}
\end{align}
Note that total exchange symmetry is preserved because the System and Label states in the second term of Eq.~(\ref{eq:dis}) are both antisymmetric.
We can now see clearly that in this case the Schur-Weyl bases provide a Schmidt decomposition of the Fock arrays, and that the completely distinguishable state has nonzero amplitude outside the totally symmetric irrep; we will discuss the generalisation of these features in Sec.~\ref{sec:UU}.

Tracing out the Label degree of freedom, we arrive at the reduced density matrices that describe the state of the System.
Another feature of the Schur-Weyl basis is that these states will be block diagonal, each block corresponding to an irrep.
Thus, ordering our triplet-singlet basis as $\left\{ \Ket{\scriptsize\young(\uparrow \uparrow)\,}, \Ket{\scriptsize\young(\uparrow \downarrow)\,}, \Ket{\scriptsize\young(\downarrow \downarrow)\,}, \Ket{\scriptsize\young(\uparrow,\downarrow)\,} \right\}$, we have
\begin{align}
\rho_\mr{i} &= \mr{Tr}_\mr{L}\left[ \ket{\psi_\mr{i}}\bra{\psi_\mr{i}} \right] = \begin{bmatrix} 0&0&0& \\ 0&1&0& \\ 0&0&0& \\ &&&0 \end{bmatrix} , \label{eq:rhomi} \\
\rho_\mr{d} &= \mr{Tr}_\mr{L}\left[ \ket{\psi_\mr{d}}\bra{\psi_\mr{d}} \right] = \frac{1}{2} \begin{bmatrix} 0&0&0& \\ 0&1&0& \\ 0&0&0& \\ &&&1 \end{bmatrix} . \label{eq:rhomd} 
\end{align}

A coincidence count occurs when both the top and bottom modes are occupied, defining the coincidence subspace spanned by $\left\{ \Ket{\scriptsize\young(\uparrow \downarrow)\,}, \Ket{\scriptsize\young(\uparrow,\downarrow)\,} \right\}$.
The projector onto this subspace has matrix representation
\begin{align}
M_{(1,1)} &= \begin{bmatrix} 0&0&0& \\ 0&1&0& \\ 0&0&0& \\ &&&1 \end{bmatrix},
\end{align}
where we have used an occupation (one excitation in each of the two System modes) in the subscript.

The unitary evolution of these input states due to the interferometer is given by the two-photon representation of the transfer matrix.
Again, in the Schur-Weyl basis this is block diagonal, specifically a direct sum of the triplet and singlet matrix representations of U$(2)$.
The matrix elements in the coincidence subspace for an arbitrary two mode interferometer with transfer matrix $U$ are
\begin{align}
U^{\otimes 2} \cong U^{\ytwo}\oplus U^{\yoneone} = \begin{bmatrix} \ast&\ast&\ast& \\ \ast&\mr{per}U&\ast& \\ \ast&\ast&\ast& \\ &&&\mr{det}U \end{bmatrix}, \label{eq:perdet}
\end{align}
where per and det are the matrix permanent and determinant functions, $\ast$ are matrix elements for events outside the coincidence subspace, and we use $\cong$ to denote the fact that $U \otimes U$ only equals $U^{\ytwo}\oplus U^{\yoneone}$ after the basis change.
This can be confirmed by direct calculation from Eq.~(\ref{eq:fund}), or equivalently by using the Schur-Weyl transformation, which for U$(2)$ is the familiar Clebsch-Gordan transformation of angular momentum theory.

The probability of a coincidence count is given by the Born rule, which from Eqs.~(\ref{eq:rhomi}--\ref{eq:perdet}) is given by
\begin{align}
P_{(1,1)} 
&= \mr{Tr}\left[ \left(U^{\ytwo}\oplus U^{\yoneone}\right)  \rho \left(U^{\ytwo}\oplus U^{\yoneone}\right)^\dag M_{(11)} \right] \label{eq:Born} \\ 
&= \mr{Tr}\left[ \left( U^{\ytwo} \rho {U^{\ytwo}}^\dag +  U^{\yoneone} \rho {U^{\yoneone}}^\dag \right) M_{(11)} \right] \\
&= \begin{cases}
 |\mr{per}U|^2 & \text{if } \rho = \rho_\mr{i} \\
 \frac{1}{2} |\mr{per}U|^2 +  \frac{1}{2} |\mr{det}U|^2 = \mr{per}|U|^2 & \text{if } \rho = \rho_\mr{d}
 \end{cases} \label{eq:squareper}
\end{align}
where we have written $|U|^2$ for the elementwise absolute value squared of a matrix $U$.

It follows that in order to see no coincidences for an indistinguishable state, which has only a triplet component, we need an interferometer whose transfer matrix permanent vanishes.
By parametrising an arbitrary $U \in \mr{U}(2)$ one can confirm that only a balanced beam splitter has this property (see Sec.~\ref{subsec:two}).
We also see that the distinguishable state has a singlet component that scatters through any $U$ according to the determinant, and since any element of U$(2)$ has $|\mr{det}U|=1$, this component will always give rise to coincidences.
Thus, in a HOM experiment one uses a balanced beam splitter to see a ``dip'' in coincidence counts in the System as one manipulates the Label degree of freedom from distinguishable to indistinguishable and back again.

\subsection{State discrimination}

By choosing to measure a coincidence count as well as $U$ to be a balanced beamsplitter, the HOM situation described above ensures that $P_{(1,1)}=0$ when the input is $\rho_\mr{i}$, while $P_{(1,1)}$ happens to be maximised when the input is $\rho_\mr{d}$ (see Sec.~\ref{subsec:two}).
This is reminiscent of what is known as unambiguous mixed state discrimination~\cite{Rudolph2003}.

A general state discrimination protocol~\cite{Bergou2004, Barnett2009} consists of two parties, a source (Alice) and a detector (Bob), who agree on an ensemble of states $\{p_k, \rho_k\}$ to be discriminated.
The source draws a random sample from this ensemble according to the distribution $\{p_k\}$ and sends it to the detector, whose task is to identify which state was sent as best as possible.
This is accomplished by finding a measurement, given by a set of POVM elements $\{E_k\}$ that maximise the expected probability of success: $\sum_k p_k \mr{Tr}[\rho_k E_k]$.
For unambiguous discrimination (UD), we have the further constraint that no mistakes are allowed to be made, that is, $\mr{Tr}[\rho_k E_j]=0$ for all $k \neq j$, at the price of having to add an outcome $E_?$ to the POVM that corresponds to failing to identify the state.

Rearranging Eq.~(\ref{eq:Born}) and defining 
\begin{align}
M_{(1,1)}(U) = \left(U^{\ytwo}\oplus U^{\yoneone}\right)^\dag M_{(1,1)} \left(U^{\ytwo}\oplus U^{\yoneone}\right) ,
\end{align}
the HOM measurement scenario described above can now be summarised by
\begin{align}
\text{find $U$ maximising}\quad & \mr{Tr}\left[ \rho_\mr{d} M_{(1,1)}(U) \right] \label{eq:HOMmax}\\ 
\text{subject to}\quad & \mr{Tr}\left[ \rho_\mr{i} M_{(1,1)}(U) \right]=0 . \label{eq:HOMcnstrt}
\end{align}
That is, find an interferometer that maximises the probability of seeing a coincidence for a distinguishable input state, subject to the constraint that it never gives coincidences for an indistinguishable input state.
It is now clear this is an instance of an UD problem, with the solution being a balanced beamsplitter in the HOM case.

This gives a direction in which to generalise the HOM scenario to any number of particles in any number of modes as a UD problem.
A key distinction from general UD is the restricted form of the available POVM elements, which must be projective measurements defined by the interferometer $U$ and the $N$-photon occupation $\underline{n}$ being detected.
In particular, we expect that known optimal measurements for two-state discrimination will not be available in linear optics.
When speaking generally about measurements we will use the notation $E$ for POVM elements, while, as above, $M_{\underline{n}} (U)$ is reserved for photon counts.
Because $M_{\underline{n}} (U)$ is degree $N$ in the variables $U$ and $U^\dag$, this measurement restriction makes the UD optimisation problem nonlinear.

\section{Background:\\ Many particles and modes}
\label{sec:manyParticles}

From the HOM example (e.g. Eq.~(\ref{eq:dis})), we see that symmetry of the states in the full System-Label space and the correlations within it play a key role in the distinguishability of the particles.
Therefore we proceed with an analysis for any arbitrary number of particles and modes using Schur-Weyl duality, and then further generalise for particles with two degrees of freedom using unitary-unitary duality~\cite{Rowe2012}.

\subsection{Schur-Weyl duality in first quantisation}
\label{subsec:SchurWeyl}

In the first quantized picture of the HOM example above, each photon was considered as a $d$-dimensional quantum system, with $d$ the total number of System and Label modes available.
Schur-Weyl duality states that the Hilbert space of $N$ qudits can be decomposed as
\begin{align}
(\mathbb{C}^d)^{\otimes N} \cong \bigoplus_{\lambda} \mathbb{C}^{\{\lambda\}} \otimes \mathbb{C}^{(\lambda)} , \label{eq:SW}
\end{align}
where $\mathbb{C}^{\{\lambda\}}$ carries irrep $\lambda$ of the group of unitary transformations on a qudit, U$(d)$, $\mathbb{C}^{(\lambda)}$ carries irrep $\lambda$ of the group of permutations of qudits, S$_N$, and $\cong$ signifies that the left and right hand sides are related by a change of basis (a Schur-Weyl transform).
Following~\cite{Harrow2005}, a Schur-Weyl basis which realises this decomposition is denoted $\ket{\lambda q p}$ where $\lambda$ labels the irrep of both the unitary and the symmetric groups simultaneously~\footnote{This one-to-one correspondence between irreps of different groups is the ``duality''.}, $q=1,2,\ldots ,d_{\{\lambda\}}$ indexes a basis of the unitary irrep, and $p=1,2,\ldots,d_{(\lambda)}$ indexes a basis of the symmetric irrep.
These dimensions can be computed, for example, by the Weyl character and hook length formulas respectively~\cite{Fulton1991}.
There is an implied dependence of $q$ and $p$ on $\lambda$, the set of which in turn depends on the number of particles $N$ and the number of modes $d$.
The irrep $\lambda = (\lambda_1, \lambda_2,\ldots,\lambda_d)$ can be specified using Young diagrams, where $\lambda_j$ is the number of boxes in row $j$, $\lambda_1 \geq \lambda_2 \cdots \geq \lambda_d$, and $\sum_j \lambda_j = N$.
The indices $q$ and $p$ correspond to the different ways of filling boxes with the numbers $\{1,\ldots, d\}$ and $\{1,\ldots, N\}$ to make semistandard (with repetition) and standard (without repetition) Young tableaux respectively, where numbers cannot decrease as you move right in a tableaux and must increase as you go down.

We can further refine this notation by observing that the basis can be chosen such that the representation theoretic \emph{weight} of a state corresponds to the occupation $\underline{n}$, which has also been called a \emph{type} in this context~\cite{Harrow2005}.
Subspaces of states with the same occupation are then invariant under the Schur-Weyl transform in this basis, and the unitary index $q$ can be uniquely specified by an occupation $\underline{n}$ and an `inner' multiplicity $r$ (the number of which is also known as a Kostka number), which accounts for the fact that there can be more than one orthogonal state with the same weight in a unitary irrep $\lambda$.
As we are focusing on the action of the unitary group, $p$ will be referred to as an `outer' multiplicity accounting for the fact that the same unitary irrep $\lambda$ can occur more than once.
We can therefore write Schur-Weyl basis states in the form $\ket{\lambda p \underline{n} r}$, where the irrep dependence of $p$, $\underline{n}$ and $r$ has again been suppressed to prevent clutter.
We will often shorten the notation such that $\ket{\lambda p \underline{n}} := \ket{\lambda, p, \underline{n}, r=1}$, $\ket{\lambda \underline{n} r} := \ket{\lambda, p=1, \underline{n}, r}$, $\ket{\lambda \underline{n}} := \ket{\lambda, p=1, \underline{n}, r=1}$, reducing clutter when the multiplicity is trivial; since $\lambda$ and $\underline{n}$ are vectors while $p$ and $r$ are scalars there should be no ambiguity.
Coincident input or output will be denoted with occupation number $\underline{1} = (1,1,\dots, 1)$, with exactly one particle in each mode, (corresponding in first quantisation to one qudit in each state).

For small $N$ and $d$, writing states in terms of Young tableaux can be more compact, as in the HOM discussion of the previous section.
The shape of a tableau is specified by $\lambda$, which is filled with mode indices specified by $\underline{n}$ following the rules for semistandard tableaux. 
The inner multiplicity $r$ corresponds to different semistandard fillings of the same $\lambda$ and $\underline{n}$, while the outer multiplicity $p$ will be labelled with a subscript.
For example, for three photons ($N=3$) in three modes ($d=3$), the coincident $\underline{n} = \underline{1} = (1,1,1)$ subspace for irrep $\lambda = (2,1) = \tiny{\yng(2,1)}$ is spanned by four states given by $p,r \in \{1,2\}$.
If we index the modes 1, 2 and 3, the two notations are related as
\begin{align}
& \ket{\lambda = (2,1), p = 1, \underline{n} = (1,1,1), r = 1} = \Ket{\scriptsize\young(12,3)_1} \\
& \ket{\lambda = (2,1), p = 1, \underline{n} = (1,1,1), r =\ 2} = \Ket{\scriptsize\young(13,2)_1} \\
& \ket{\lambda = (2,1), p = 2, \underline{n} = (1,1,1), r = 1} = \Ket{\scriptsize\young(12,3)_2} \\
& \ket{\lambda = (2,1), p = 2, \underline{n} = (1,1,1), r = 2} = \Ket{\scriptsize\young(13,2)_2},
\end{align}
while, e.g., the $\underline{n}=(2,1,0)$ subspace for irrep $\lambda = (2,1)$ is spanned by the two states
\begin{align}
& \ket{\lambda = (2,1), p = 1, \underline{n} = (2,1,0), r = 1} = \Ket{\scriptsize\young(11,2)_1} \\
& \ket{\lambda = (2,1), p = 2, \underline{n} = (2,1,0), r = 1} = \Ket{\scriptsize\young(11,2)_2},
\end{align}
because the Young tableau $\scriptsize\young(12,1)$ is not semistandard and therefore such states do not exist.

\subsubsection{Implementation of the Schur-Weyl transform}
\label{subsec:SchurWeylImplement}
An example of a Schur-Weyl transformation is the triplet-singlet basis change given in Eqs.~(\ref{eq:11} - \ref{eq:00}), where (when $d=2$) it is the same as the well known Clebsch-Gordan transformation.
There are several ways to implement this basis change more generally~\cite{Gelfand1950,Moshinsky1963}; we use the method described in Ref.~\cite{Rowe2012}, which we will briefly outline here.

Every irrep $\{\lambda\}$ of U$(d)$ can be assigned a highest weight state, which is annihilated by an appropriate set of raising operators that are realised in terms of the bosonic creators and annihilators.
Given as a Young tableau, this state can be expressed in terms of single particle (qudit) states using Slater determinants; the single particle basis is indexed by the $d$ modes.
In much the same way as is done for U$(2)$ in angular momentum theory, we then use corresponding lowering operators to find a set of states that span the irrep.
The size of this set is known, namely $d_{\{\lambda\}}$.
A Gram-Schmidt procedure is then used to orthonormalise the set, (note that there is freedom in choosing how to do so when there are multiplicities, see e.g. Sec.~\ref{subsec:singlyD}).
Outer multiplicities are handled by utilising the dual S$_N$ action to permute a highest weight state in order to find corresponding highest weights for the multiple copies of irrep $\{\lambda\}$.
Again, the number of these states is known, namely $d_{(\lambda)}$, and orthonormalisation is required.
The lowering procedure is then repeated until a complete set of $\lambda$ states are found.
Iterating through all $\lambda$ then gives a complete set of states $\{\ket{\lambda q p}\}$, from which we can determine the required basis transformation.
Transformations for different $N$ and $d$ can be computed once and stored for later use.

\subsection{Unitary-unitary duality}
\label{sec:UU}
In the HOM example we saw that each photon had two degrees of freedom, the System and the Label, and that, as bosons, first quantised multiphoton states had to be totally symmetric under particle permutations.
Independently decomposing both the System and Label Hilbert spaces according to Schur-Weyl, one is then led to ask what states of the form
\begin{align}
\sum_{\substack{\lambda q p\\ \lambda' q' p'}} \psi^{\lambda q p}_{\lambda' q' p' } \ket{\lambda q p}_\mr{S} \ket{\lambda' q' p'}_\mr{L}
\end{align}
are totally symmetric?
This can be viewed as a coupling problem for irreps of the symmetric group -- we wish to construct composite states of `permutational momentum zero'.
The answer turns out much like it does in angular momentum theory: that $\lambda$, $p$ must equal $\lambda'$, $p'$, respectively, and that the coupling coefficients are all equal and independent of $p$~\cite{Hamermesh1962,Adamson2008}. 
Thus totally symmetric pure System-Label states are of the form
\begin{align}\label{eq:symSL}
\sum_{\lambda q q'} \psi_{\lambda q q' } \ket{\lambda q q'}_\mr{SL} ,
\end{align}
where we have defined
\begin{align}\label{eq:lambdasymSL}
\ket{\lambda q q'}_\mr{SL}
 &:= \frac{1}{\sqrt{d_{(\lambda)}}} \sum_{p=1}^{d_{(\lambda)}} \ket{\lambda q p}_\mr{S} \ket{\lambda q' p}_\mr{L} .
\end{align}
These states carry the symmetric irrep of the `global' unitary group, $\mathrm{U}(d_\mr{S} d_\mr{L})$, acting on the $d_\mr{S} d_\mr{L}$ modes of the combined System and Label.
As discussed above, we can replace $q$ with pairs $\underline{n},r$ in all of these expressions.

Equations~(\ref{eq:symSL}, \ref{eq:lambdasymSL}) imply a decomposition of the totally symmetric irrep of U$(d_\mr{S} d_\mr{L})$ into irreps of its unitary subgroups $\mathrm{U}(d_\mr{S})$ and $\mathrm{U}(d_\mr{L})$ that act on the System and Label independently.
These irreps are labelled simultaneously by $\lambda$, hence ``unitary-unitary duality'':
\begin{align}
\mathrm{Sym}\left( (\mathbb{C}^{d_\mr{S}} \otimes \mathbb{C}^{d_\mr{L}})^{\otimes N} \right) \cong \bigoplus_{\lambda} \mathbb{C}^{\{\lambda\}_\mr{S}} \otimes \mathbb{C}^{\{\lambda\}_\mr{L}} ,
\end{align}
where we include subscripts on the right hand side to remind us which unitary subgroups the irreps belong to~\footnote{An analogous result holds for fermions and the antisymmetric subspace, where one couples irreps with their transpose Young diagrams.}.
An interferometer $U$ is given by an element of the System unitary subgroup U$(d_\mr{S})$, and thus it acts on states in irrep $\lambda$ according to the irreducible matrix representation $U^\lambda$
\begin{align}
U : \ket{\lambda q q'}_\mr{SL} \mapsto \sum_{q''} \ket{\lambda q'' q'}_\mr{SL} U^\lambda_{q''q} .
\end{align}

Just as with a single degree of freedom, the space of second quantized $d_\mr{S} \times d_\mr{L}$ Fock arrays can be put into one-to-one correspondence with first quantized totally symmetric states by the procedure exemplified in Eqs.~(\ref{eq:2ndind} - \ref{eq:1stind}).
Thus we can write an arbitrary partially distinguishable state, which is an element of the totally symmetric subspace of $(\mathbb{C}^{d_\mr{S}} \otimes \mathbb{C}^{d_\mr{L}})^{\otimes N}$, in a basis of first quantized states given by Eq.~(\ref{eq:lambdasymSL}).
We may now trace out the Label to arrive at mixed states describing any partially distinguishable state of $N$ photons in $d_\mr{S}$ modes.
We can order the basis so that the reduced System state and the action of any System interferometer will both be block diagonal according to irreps $\lambda$, a potentially significant simplification.

\subsection{States of interest}
\label{sec:states}
We will focus our attention on three types of $N$-photon states: completely indistinguishable, singly distinguishable, and completely distinguishable, described below (the general case will be discussed in Sec.~\ref{sec:discuss}).
We are not considering loss (where entire qudits would be traced out), so $N$ will be fixed throughout.
Situations with mixed System-Label states and partial distinguishability can be written in terms of the basis~\cite{Rohde2012};
we give examples of this generality with partial distinguishability for two photons in two modes in Sec.~\ref{subsec:twotwo}, and of mixed System-Label states for three photons in three modes in Sec.~\ref{subsec:mixedthree}.
Otherwise we will restrict ourselves to the case where the total System-Label state is pure, corresponding to a source that produces states that are always (in)distinguishable in exactly the same way; generalization is, in principle, straightforward.

The most distinguishable $N$ photons can be is for each Label to be in an orthogonal state, and so $d_\mr{L} \leq N$.
In practice the Label space could be much larger, but in order to describe the particles' distinguishability we need only consider the subspace spanned by the Label states, which can be at most $N$ dimensional.
In order to set $d_\mr{S}$, consider first two photons who share the same state of either degree of freedom; obviously that state is symmetric, and so in order to maintain total symmetry -- or by unitary-unitary duality -- the state of the other degree of freedom must also be symmetric, cf. Eq.~(\ref{eq:1stind}).
This restricts the combined state to a subspace of those allowed in Eq.~(\ref{eq:lambdasymSL}), and so is not completely general.
This argument extends to any number of photons, thus to consider arbitrary distinguishability we must have input states that have a single photon in each System mode, thus $d_\mr{S} \geq N$.
Unless indicated otherwise, we will consider the case with $d_\mr{S}=d_\mr{L}=N$.
The reader may wish to refer ahead to Sec.~\ref{subsec:three} for concrete examples of the following.

\subsubsection{Completely indistinguishable states}
\label{subsec:compI}
A completely indistinguishable state is one in which every photons' Label state is the same.
As mentioned above, such a state lies in the symmetric Label subspace with $\lambda=(N)$.
Since the symmetric irrep of S$_N$ is one dimensional, $d_{(N)}=1$ and Schur-Weyl duality tells us that the corresponding unitary irrep is always outer multiplicity free.
Moreover, $(N)$ is also inner multiplicity free, (there is only one way to symmetrise a product of single particle states), so we can replace $q$ with the System occupation $\underline{1}$, and $q'$ with the Label occupation $(N,\underline{0})$ (ordering our Label modes such that the occupied one is first, and with the understanding that the list of zeroes is as long as it needs to be, in this case $N-1$).
The total state in Eq.~(\ref{eq:lambdasymSL}) therefore becomes
\begin{align}
\hat{a}^\dag_{11}\hat{a}^\dag_{21}\cdots\hat{a}^\dag_{N1}\ket{\mr{vac}}
&={\tiny \Ket{\setlength\arraycolsep{2pt}\renewcommand{\arraystretch}{0.75}\begin{matrix} 1&0&\cdots&0\\1&0&\cdots&0\\\vdots&&\ddots&\\1&0&\cdots&0 \end{matrix}}} \\
&=\mr{Sym}\left(\ket{11}\ket{21}\cdots\ket{N1}\right) \\
&=\ket{(N),\underline{1}}_\mr{S}\ket{(N),(N,\underline{0})}_\mr{L} ,
\end{align}
where we have included $N-1$ redundant zero columns in the Fock array so we can easily compare with the other states in this section.
In the second line we have written the state in the single particle basis, cf. Eqs.~(\ref{eq:Symi}, \ref{eq:Symd}), and we have suppressed trivial multiplicities in the last line.
We see that this is always a product state, with no correlation between the System and Label, as expected for completely indistinguishable particles.
The reduced System state is
\begin{align}
\rho_\mr{i}
&= \mr{Tr}_\mr{L}\big[ \ket{(N),\underline{1}}\ket{(N),(N,\underline{0})}\bra{(N),\underline{1}}\bra{(N),(N,\underline{0})} \big] \\
&= \ket{(N),\underline{1}}\bra{(N),\underline{1}} , \label{eq:rhoi}
\end{align}
supported on the one dimensional intersection of the symmetric System subspace given by $(N)$ with the coincident subspace defined by the System occupation number $\underline{1}$.

\subsubsection{Singly distinguishable states}
\label{subsec:singlyD}
The next state we consider is one where a single photon has become distinguishable from the rest; assuming all efforts are being made to produce the completely indistinguishable state, this should be the most likely error to occur.
Ordering our modes so that the `bad' photon is in System mode $N$ and Label mode 2, we have
\begin{align}
\hat{a}^\dag_{11}\hat{a}^\dag_{21}\cdots\hat{a}^\dag_{N2}\ket{\mr{vac}}
&={\tiny \Ket{\setlength\arraycolsep{2pt}\renewcommand{\arraystretch}{0.75}\begin{matrix} 1&0&\cdots&0\\1&0&\cdots&0\\\vdots&&\ddots&\\0&1&\cdots&0 \end{matrix}}} \\
&=\mr{Sym}\left(\ket{11}\ket{21}\cdots\ket{N2}\right), \label{eq:NSym}
\end{align}
where in the last line we have not yet performed the Schur-Weyl transform.
Considering this symmetrisation, one observes that although all $N!$ permutations of the $N$ distinct System indices will occur, since only two distinct Label modes are involved there are only $N$ single particle states available to the Label degree of freedom, namely those with the $j^\mr{th}$ photon in Label mode 2 and the rest in Label mode 1; denote these states $\ket{2_j}_\mr{L}$.
Such a Label state will be perfectly correlated to all System states with the $j^\mr{th}$ photon in mode $N$; for each $j$ we can factor these $(N-1)!$ System states off, and denote the resulting normalised state $\ket{N_j}_\mr{S}$.
Thus in the System-Label basis, the singly distinguishable state can be written as
\begin{align}\label{eq:SLsingle}
\mr{Sym}\left(\ket{11}\ket{21}\cdots\ket{N2}\right)
&=
\frac{1}{\sqrt{N}} \sum_{j=1}^N \ket{N_j}_\mr{S} \ket{2_j}_\mr{L} ,
\end{align}
e.g. Eqs.~(\ref{eq:1stind}, \ref{eq:1stdis}).
These sets of states are orthonormal, and we recognise this as an entangled state with Schmidt coefficients $1/\sqrt{N}$.

Now consider Schur-Weyl transforming this state into the form of Eq.~(\ref{eq:symSL}).
Because there are only two distinct Label modes involved, the only Label irreps that can occur are those whose Young diagrams have two or fewer rows.
Moreover, because only a single photon is `bad', the only two rowed diagram allowed is that with a single box in the second row.
Thus the Label state is supported only by irreps $\lambda=(N)$ and $(N-1,1)$.
By unitary-unitary duality, the System is therefore also supported only on these two irreps.
The totally symmetric irrep $(N)$ is always both inner and outer multiplicity free; for irrep $(N-1,1)$, the outer multiplicity  is $d_{((N-1,1))} = N-1$.
It remains only to work out the inner multiplicities for irrep $(N-1,1)$.
The System and Label have occupations $\underline{1}$ and $(N-1,1,\underline{0})$ respectively (the marginals of the Fock array).
There is only one Young tableau of shape $(N-1,1)$ consistent with occupation $(N-1,1,\underline{0})$, (that with the 2 in the second row box), so the Label states are inner multiplicity free.
The System occupation $\underline{1}$ is consistent with $N-1$ Young tableau of shape $(N-1,1)$, (all those without an $N$ in the second row box), and so the System inner multiplicity is $N-1$.
Inserting these observations into Eq.~(\ref{eq:symSL}), the Schur-Weyl transformed state is
\begin{align}
& \psi_{(N),\underline{1},1,(N-1,1,\underline{0}),1} \ket{(N),1,\underline{1},1}_\mr{S}
\ket{(N),1,(N-1,1,\underline{0}),1}_\mr{L} \nonumber\\
&+
\sum_{r=1}^{N-1} \frac{\psi_{(N-1,1),\underline{1},r,(N-1,1,\underline{0}),1}}{\sqrt{N-1}} \sum_{p=1}^{N-1}
\ket{(N-1,1),p,\underline{1},r}_\mr{S} \nonumber\\
&\quad\qquad\qquad\times \ket{(N-1,1),p,(N-1,1,\underline{0}),1}_\mr{L} .
\end{align}
We can factor the second term and redefine coefficients to yield another Schmidt decomposition:
\begin{align}\label{eq:obegg}
& \psi_{(N)} \ket{(N),1,\underline{1},1}_\mr{S}
\ket{(N),1,(N-1,1,\underline{0}),1}_\mr{L} \nonumber\\
&+
\frac{\psi_{(N-1,1)}}{\sqrt{N-1}}\sum_{p=1}^{N-1}
\left(\sum_{r=1}^{N-1}\phi_r\ket{(N-1,1),p,\underline{1},r}_\mr{S}\right) \nonumber\\
&\quad\qquad\qquad\times \ket{(N-1,1),p,(N-1,1,\underline{0}),1}_\mr{L} .
\end{align}
Because the Schur-Weyl transformations yielding Eq.~(\ref{eq:lambdasymSL}) are performed independently, the System-Label entanglement cannot be changed.
From Eq.~(\ref{eq:SLsingle}) we know that the Schmidt coefficients are all $1/\sqrt{N}$, so we must have $\psi_{(N)}=1/\sqrt{N}$ and $\psi_{(N-1,1)}=\sqrt{(N-1)/N}$.
The amplitudes $\phi_r$ do not affect this entanglement at all -- they depend on how one chooses to orthonormalise multiplicities in the Schur-Weyl transform, and encode the fact that we chose the `bad' photon to be in System mode $N$.
We can always choose $r=1$ to correspond to this specific situation, and then use the subgroup of U$(d_\mr{S})$ that permutes System modes to find the states corresponding to the `bad' photon being in any other mode.

Making this choice and tracing out the Label in Eq.~(\ref{eq:obegg}) yields the singly distinguishable reduced state (now suppressing trivial multiplicities)
\begin{align}
\rho_{\mr{s}}
=&
\frac{1}{N} \ket{(N),\underline{1}}\bra{(N),\underline{1}} \nonumber\\ 
&+\frac{1}{N} \sum_{p=1}^{N-1} \ket{(N-1,1),p,\underline{1},1}\bra{(N-1,1),p,\underline{1},1} .
\label{eq:rhos}
\end{align}
We see that this is mixed over $N$ dimensions of the coincident subspace, overlapping the symmetric and `almost symmetric' $(N-1,1)$ irreps.

\subsubsection{Completely distinguishable states}
\label{subsec:compD}
A completely distinguishable state has each particle in a distinct Label mode, paired with a unique System mode.
We can choose to order the modes such that the corresponding Fock array is diagonal, cf. Eq.~(\ref{eq:d}). 
Generalising the symmetrisation procedure of Eqs.~(\ref{eq:2ndind} - \ref{eq:1stind}) to $N$ particles, one finds that all $N!$ possible terms will occur in the single particle picture, and they will each occur once.
The unique pairing of System and Label modes manifests as maximal entanglement between the System and Label single particle states in the coincident subspace. 
As above, because the Schur-Weyl transformations yielding Eq.~(\ref{eq:lambdasymSL}) are performed independently, the System-Label entanglement is preserved.
This means that the transformed state must also be maximally entangled with the same Schmidt rank.
Thus
\begin{align}
\hat{a}^\dag_{11}\hat{a}^\dag_{22}\cdots\hat{a}^\dag_{NN}\ket{\mr{vac}}
&={\tiny \Ket{\setlength\arraycolsep{2pt}\renewcommand{\arraystretch}{0.75}\begin{matrix} 1&0&\cdots&0\\0&1&\cdots&0\\\vdots&&\ddots&\\0&0&\cdots&1 \end{matrix}}} \\
&=\mr{Sym}\left(\ket{11}\ket{22}\cdots\ket{NN}\right) \\
&=\frac{1}{\sqrt{N!}} \sum_{\lambda p r} \ket{\lambda, p, \underline{1}, r}_\mr{S} \ket{\lambda, p, \underline{1}, r}_\mr{L} ,
\end{align}
with the sum running over all allowed values of irrep, outer, and inner multiplicities.
The completely distinguishable reduced System state is therefore
\begin{align}
\rho_\mr{d}
=& \frac{1}{N!} \sum_{\lambda p r} \ket{\lambda,p,\underline{1},r}\bra{\lambda,p,\underline{1},r} \label{eq:1rhod} \\
=& \frac{1}{N!} \ket{(N),\underline{1}}\bra{(N),\underline{1}} \nonumber\\
&+ \frac{1}{N!} \sum_{\lambda\neq (N), p, r} \ket{\lambda,p,\underline{1},r}\bra{\lambda,p,\underline{1},r} , 
\label{eq:rhod}
\end{align}
which is completely mixed over the $N!$ dimensional coincident subspace.

\subsection{Unitary parametrisation}
\label{subsec:Reck}
\begin{figure}[h]
\includegraphics[width=0.5\textwidth]{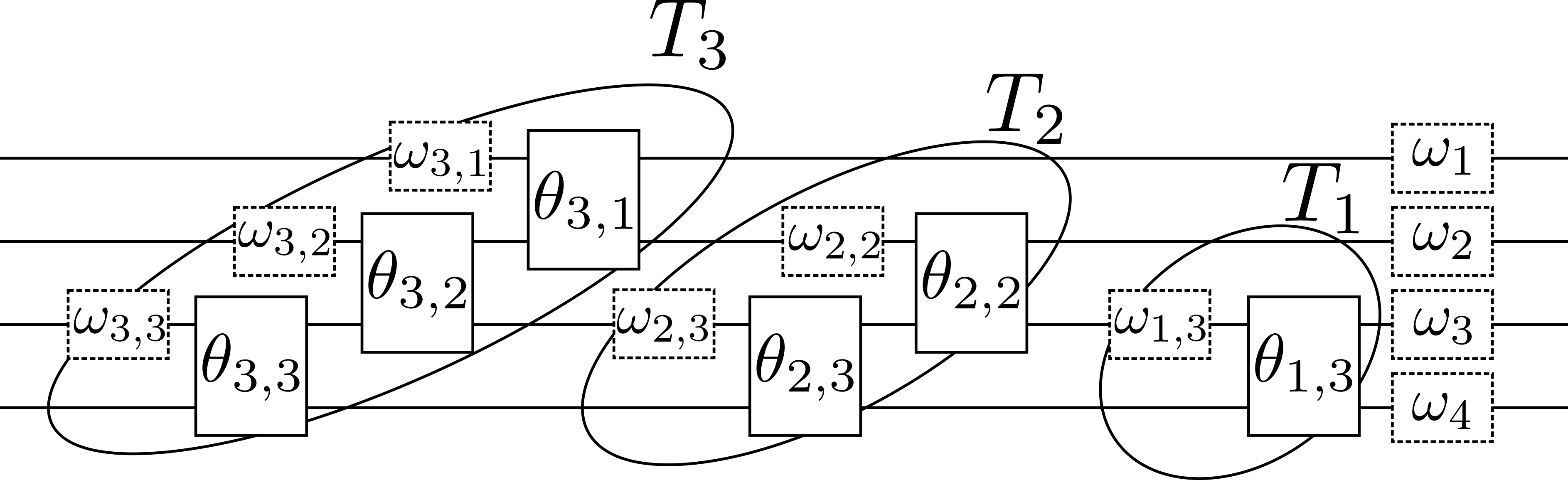}
\caption{
Example of a Reck scheme parametrising an arbitrary unitary transformation on four modes ($d_\mr{S}=4$), grouped into `layers' $T_k$.
Each one- (phaseshifter) and two-mode (beamsplitter) subtransformation contributes one real parameter.
Only the phaseshifters situated between beamsplitters ($\omega_{1,3}, \omega_{2,2}, \omega_{2,3}$) contribute to our problem.
}
\label{fig:reck}
\end{figure}

The unitary subgroup U$(d_\mr{S})$ corresponds to the set of interferometers that act on the System modes.
We can parametrise these unitaries with what is known as a Reck scheme in optics~\cite{Hurwitz1897, Reck1994}, decomposing an arbitrary $U$ into a sequence of single mode unitaries (phaseshifters) and unitaries that act on neighbouring modes (beamsplitters).
As shown in Fig.~\ref{fig:reck}, such a scheme can be viewed as $d_\mr{s}-1$ layers, indexed by $k$, each with $k$ phaseshifters and beamsplitters, followed by a final phase shift on each mode.
Because we are only interested in number state inputs and number counting measurements, only the phaseshifters between beamsplitters play a role.
Hereafter when we refer to $U$ we will therefore be referring to this smaller interferometer, without the initial and final sets of phaseshifters.

\subsection{Measurements}
\label{subsec:measurements}

We will assume that we have access to photon number resolving detectors for the System (see Sec.~\ref{sec:discuss} for a discussion of a relaxation).
The measurement POVM elements are projections on all states with photon occupation $\underline{n}$, $M_{\underline{n}} = \sum_{\lambda p r} \ket{\lambda p \underline{n} r} \bra{\lambda p \underline{n} r}$.
Note that this includes projections onto System states that are not symmetric; as shown in Eq.~(\ref{eq:dis}), distinguishable states can contain non-symmetric System components that still give rise to clicks.
Comparing with Eq.~(\ref{eq:1rhod}), we see that $M_{\underline{1}} = N!\,\rho_\mr{d}$ -- that is, up to normalisation, a coincidence count is a projection onto the completely distinguishable state.
As discussed above, we will usually include the interferometer in our definition of a measurement, yielding parametrised POVM elements 
\begin{equation}
M_{\underline{n}} (U) = \left(\oplus_\lambda U^\lambda\otimes\Id^{\lambda} \right)^\dag M_{\underline{n}} \left(\oplus_{\lambda'} U^{\lambda'}\otimes\Id^{\lambda'} \right) ,
\label{eq:photon_counting}
\end{equation}
where $\Id^{\lambda}$ corresponds to the irrep of the identity permutation in accordance with Eq.~(\ref{eq:SW}), (note that we omit this when it is only one dimensional, e.g. Eq.~(\ref{eq:perdet}) ).

\section{Discrimination of distinguishable states}
\label{sec:discrimination}

We will be interested in two problems: discriminating the completely indistinguishable state, $\rho_\mr{i}$, from the distinguishable states (i) $\rho_\mr{s}$ and (ii) $\rho_\mr{d}$.
From Eqs.~(\ref{eq:rhoi}, \ref{eq:rhos}, \ref{eq:rhod}), we observe that each of these states is of the form
\begin{equation}
\rho= \alpha \rho_\mr{i} + (1-\alpha) \rho_{\bar{\mr{i}}}, \quad \alpha\neq 0, \label{eq:statesetup}
\end{equation} 
where $\rho_\mr{i}$ is pure, and $\rho_{\bar{\mr{i}}}$ is diagonal in the Schur-Weyl basis with support outside the symmetric subspace $\lambda=(N)$.
From well known results for the discrimination of two mixed states~\cite{Herzog2004}, the fact that $\rho_\mr{i}$ lies within the support of the mixed state to be discriminated means that the optimal measurement is essentially the same for either Minimum Error or Unambiguous Discrimination; one wishes to project onto the support of $\rho_{\bar{\mr{i}}}$. 
In particular for UD, the error-free constraint means that 
we are forced to set $E_\mr{i}=0$, and thus the prior probabilities do not affect the optimal choice of measurement operators. 
This reflects the fact that there is no way to unambiguously discriminate the indistinguishable state $\rho_\mr{i}$ -- we can either conclude that the state was distinguishable by observing an output that is completely suppressed by quantum interference, or fail to conclude anything at all.
Our task is therefore to minimise the probability of failure $E_? = \Id - E_{\mr{s},\mr{d}}$, equivalently maximising the probability of unambiguously detecting a singly or completely distinguishable state, respectively.

If our measurements are unrestricted, the best choice of POVM is to project onto the nonsymmetric subspace.
This choice is suitable for not only the states $\rho_\mr{s,d}$, but by extension any state to be discriminated from $\rho_\mr{i}$.
However, as mentioned in Sec.~\ref{subsec:measurements}, in practice we only have access to number counting measurements -- we will therefore want to approximate this projection as best possible.
The approximation will be sensitive to the state we are discriminating: for example, Eqs.~(\ref{eq:rhos}, \ref{eq:rhod}) show that $\rho_\mr{s}$ can be optimally discriminated by projecting onto only the $(N-1, 1)$ irrep, while for $\rho_\mr{d}$ one wants to project on to all of the nonsymmetric irreps.
As we will see, this can lead to different interferometers being optimal for discriminating different distinguishable states in linear optics.

\subsection{Restriction to linear optical measurements}
\label{subsec:discrimination}
In order to discriminate distinguishability in linear optics we wish to find the best we can do with the measurements we have, namely those in Eq.~(\ref{eq:photon_counting}).
In the HOM case, the UD problem described by Eqs.~(\ref{eq:HOMmax}, \ref{eq:HOMcnstrt}) involves only a single occupation POVM element, the coincidence count $M_{\underline{n}}(U)$ with $\underline{n}=(1,1)$.
There are many ways we can approach the generalization of the HOM case.
One way would be, given a set of measurement operators, for each $\underline{n}$ we find $U$ maximising $\Tr \left[\rho M_{\underline{n}} (U) \right]$ subject to $ \Tr \left[ \rho_{\mr{i}} M_{\underline{n}} (U) \right]=0$.
Notice that any $\underline{n}$ that can be made to satisfy $ \Tr \left[ \rho_{\mr{i}} M_{\underline{n}} (U) \right]=0$ for a suitable $U$ is an unambiguous discriminator, but is not necessarily the optimal choice.
In general, it is possible for multiple occupations $\underline{n}$ to satisfy the UD constraint simultaneously, contributing to the probability of success.

We therefore wish to find the subset of all discriminating occupations, call it $D$, that optimises the success probability for the same choice of $U$:
\begin{align}
\text{find $U$ and $D$ maximising}\quad & \sum_{\underline{n} \in D} \Tr \left[\rho M_{\underline{n}} (U) \right] \label{eq:Allmax}\\ 
\text{subject to, for all }  \underline{n} \in D,\quad & \Tr \left[ \rho_{\mr{i}} M_{\underline{n}} (U) \right]=0 . \label{eq:Allcnstrt}
\end{align}
Note that the quantity we are maximizing gives us the total probability of successful discrimination, which is the sum over all the unambiguously discriminating events in the set of occupations $D$.

While the first optimisation focuses on giving an optimal interferometer for discrimination given a specific measurement pattern, the second optimisation focuses on the highest probability of discrimination across all measurement patterns.
In general we find that these two problems give different optimal interferometers; here we will focus on the latter `complete' optimisation over both $U$ and $D$, see Sec.~\ref{sec:discuss} for a discussion of a variation of the problem.

\subsection{Scattering probabilities}
\label{subsec:simp}
Let us look at what the probability of a specific measurement pattern $\underline{n}$ being detected at the output of an arbitrary interferometer $U$ is for the states of interest, starting with the completely distinguishable state.
From Eqs.~(\ref{eq:rhod}) and (\ref{eq:photon_counting}),
\begin{widetext}
\begin{align}
\Tr \left[ \rho_{\mr{d}} M_{\underline{n}}(U) \right]
& = \Tr \left[ \left( \frac{1}{N!} \sum_{\lambda, p, r} \ket{ \lambda, p, \underline{1}, r} \bra{ \lambda, p, \underline{1}, r}\right) \left(\oplus_\mu U^\mu\otimes\Id^\mu \right)^\dag \left(\sum_{\lambda', p', r'} \ket{ \lambda', p', \underline{n}, r'} \bra{ \lambda', p', \underline{n}, r'}\right)\left(\oplus_{\mu'} U^{\mu'}\otimes\Id^{\mu'} \right) \right] \nonumber \\
& =  \frac{1}{N!}  \sum_{\lambda , p, r, r'}  \Tr \left[ \ket{ \lambda, p,  \underline{n}, r} \bra{ \lambda, p, \underline{n}, r} \left( U^\lambda\otimes\Id^\lambda \right)  \ket{ \lambda, p, \underline{1}, r'} \bra{ \lambda, p,  \underline{1}, r'} \left( U^\lambda\otimes\Id^\lambda \right)^\dag \right]  \nonumber  \\
& =  \frac{1}{N!}   \sum_{\lambda , p, r, r'}  | \bra{ \lambda, p, \underline{n},r} U^{\lambda}\otimes\Id^\lambda  \ket{ \lambda, p, \underline{1}, r'} |^2 \nonumber \\ 
& = \frac{1}{N!} \, \sum_{\lambda , r, r'} d_{(\lambda)} | \bra{ \lambda, \underline{n},r} U^{\lambda}  \ket{ \lambda, \underline{1}, r'} |^2 , \label{eq:simplifiedFD}
\end{align}
\end{widetext}
where in the last line we have used the fact that outer multiplicities $p$ give rise to identical copies of unitary irreps to write the probability in terms of irreducible unitary matrix elements.
When $r=r'=1$ these matrix elements are immanants~\cite{Littlewood1934} of a matrix $U_{\underline{1}}^{\underline{n}}$ whose rows and columns are determined by the input and output occupations of the interferometer given by $U$~\cite{Kostant1995,deGuise2016}.
Moreover, the completely distinguishable case can be interpreted as independent classical particles evolving stochastically~\cite{Aaronson2013}, leading to the remarkable fact that the sum in Eq.~(\ref{eq:simplifiedFD}) can always be written in terms of the permanent of the matrix given by the elementwise square amplitudes of $U_{\underline{1}}^{\underline{n}}$, cf. Eq.~(\ref{eq:squareper}) and note that $U_{\underline{1}}^{\underline{1}}=U$.

The calculation for the singly distinguishable and completely indistinguishable state is the same as Eq.~(\ref{eq:simplifiedFD}), only with fewer irreps occurring.
Recalling from Sec.~\ref{subsec:singlyD} that $d_{((N-1,1))}=N-1$, Eq.~(\ref{eq:rhos}) gives
\begin{align}
&\Tr \left[ \rho_{\mr{s}} M_{\underline{n}} (U) \right] = \frac{1}{N} | \bra{ (N), \underline{n}} U^{(N)}  \ket{ (N),  \underline{1}} |^2 \nonumber \\
& + \frac{N-1}{N} \sum_{r}  |\bra{ (N-1,1), \underline{n}, r} U^{(N-1,1)}  \ket{  (N-1,1), \underline{1},1} |^2 , \label{eq:simplifiedOBE}
\end{align}
where the sum is over all $r$ consistent with $\underline{n}$, and Eq.~(\ref{eq:rhoi}) gives
\begin{align}
&\Tr \left[ \rho_{\mr{i}} M_{\underline{n}} (U) \right] = |\bra{ (N), \underline{n}} U^{(N)} \ket{ (N), \underline{1}}|^2 ,\label{eq:simplifiedID}
\end{align}
where as mentioned above these matrix elements are expressible in terms of $\mr{per} U_{\underline{1}}^{\underline{n}}$~\cite{Bhatia1997}.

We observe that not all occupations are useful for unambiguous discrimination.
Measurements where all the photons are bunched into a single mode only occur in the symmetric irrep, that is, if $\underline{n} = (0,..,0,N,0,...,0)$, then $M_{\underline{n}} =  \ket{(N), 1, \underline{n},1} \bra{(N), 1, \underline{n}, 1}$.
In this case Eqs.~(\ref{eq:simplifiedFD}) and (\ref{eq:simplifiedOBE}) are proportional to Eq.~(\ref{eq:simplifiedID}), and since Eq.~(\ref{eq:Allcnstrt}) has to be satisfied, they will always give zero.
Completely bunched events can therefore never help discriminate the indistinguishable state, and we will exclude such events from our searches.

\subsection{Numerical optimisation approach}
\label{subsec:numerical}

In the Results section there is a mixture of analytical and numerical results.
To construct the cost function for our numerical work we took into consideration the following criteria: the measurement operator $M_{\underline{n}}$ can only be included in the optimisation if Eq.~(\ref{eq:Allcnstrt}) is satisfied; when this is the case it is added to a sum being optimised as per Eq.~(\ref{eq:Allmax}).
The cost function chosen was
\begin{align}
C(U) = - \sum_{\underline{n}} \exp \left(- \xi \Tr \left[ \rho_{\mr{i}} M_{\underline{n}}(U) \right] \right) \Tr \left[ \rho M_{\underline{n}}(U) \right] , \label{eq:costfn}
\end{align}
where $\xi$ is adjusted (usually depending on the choice of $N$, and ranging from $2$ to $60$) to penalise results where $M_{\underline{n}}$ might be added to Eq.~(\ref{eq:Allmax}) and optimised without satisfying Eq.~(\ref{eq:Allcnstrt}).
A high penalty $\xi$ guarantees that the value of $\Tr\left[\rho_{\mr{i}} M_{\underline{n}}(U)\right]$ is close to zero before $\Tr \left[ \rho M_{\underline{n}}(U) \right]$ is optimised and added to the sum.
Combining this with the Eqs.~(\ref{eq:simplifiedFD}) and (\ref{eq:simplifiedOBE}) we have 

\begin{align}
C_{\mr{d}}(U)
=& \frac{-1}{N!} \sum_{\lambda \neq (N)} d_{(\lambda)}\sum_{\underline{n}} e^{- \xi |\bra{ (N), \underline{n}} U^{(N)} \ket{ (N), \underline{1}}|^2} \nonumber\\
&\times \sum_{r, r'}  | \bra{ \lambda, \underline{n},r}  U^{\lambda}  \ket{ \lambda, \underline{1}, r'} |^2  , \label{eq:costfnFD} \\
C_{\mr{s}}(U) 
=&  \frac{1-N}{N} \sum_{\underline{n}, r} e^{- \xi |\bra{ (N), \underline{n}} U^{(N)} \ket{ (N), \underline{1}}|^2 } \nonumber\\
&\times | \bra{ (N-1,1), \underline{n}, r} U^{(N-1,1)}  \ket{  (N-1,1), \underline{1}} |^2. \label{eq:costfnOBE}
\end{align}

Python was used to optimise these functions with the scipy library function \texttt{basinhopping} using Broyden–Fletcher–Goldfarb–Shanno (BFGS) as the optimisation algorithm.
The seeds were generated using numpy random number generation.
Though this optimisation function will help us explore the space and reach fairly close to the global minimum, it can neither guarantee that minimum is global, nor does it exactly solve the original optimisation problem.
This will be problematic with minima that are close together, as for example $\exp \left(- \xi \Tr\left[\rho_{\mr{i}} M_{\underline{n}}(U)\right] \right)$ gets closer to $1$ for values of $\Tr\left[\rho_{\mr{i}} M_{\underline{n}}(U)\right]$ that are close to $0$.
In some situations this value can be quite high combined with a high value of  $\Tr \left[\rho M_{\underline{n}}(U)\right]$, skewing the results towards a possible non-optimal solution for the original problem.
We could avoid this by choosing an appropriately high $\xi$ as a function of the number of occupations ${N+d_\mr{S}-1 \choose N}$, however, if too high, $\exp \left(- \xi \Tr \left[\rho_{\mr{i}} M_{\underline{n}}(U) \right] \right)$ will behave like a step function, which does not reward transitional values enough.
Therefore, we do not make any strong claims of optimality for the interferometers found numerically when they do not saturate the general bounds presented in Sec.~\ref{subsec:bounds}.

\section{Results}

\subsection{General bounds}
\label{subsec:bounds}

Recall from Sec.~\ref{sec:discrimination} the best possible unrestricted discrimination measurement is to project onto the nonsymmetric subspace, $E_{\overline{(N)}} = \sum_{\lambda \neq (N), p, \underline{n}, r} \ket{\lambda p \underline{n} r} \bra{\lambda p \underline{n} r}$.
Such a POVM element would be equally good for both singly and completely distinguishable states, and indeed any distinguishable state of the form in Eq.~(\ref{eq:statesetup}).
The success probability of such a measurement is given by
\begin{align}
\Tr & \left[ \rho \left(\oplus_\lambda U^\lambda\otimes\Id^\lambda \right)^\dag E_{\overline{(N)}} \left(\oplus_\lambda U^\lambda\otimes\Id^\lambda \right) \right] \nonumber \\
&=\Tr \left[\left(\alpha\rho_\mr{i}+(1-\alpha)\rho_{\bar{\mr{i}}}\right) E_{\overline{(N)}} \right] \nonumber \\
&= 1 - \alpha \nonumber \\
&=\begin{cases}
 1-\frac{1}{N} & \text{if } \rho = \rho_\mr{s} \\
 1-\frac{1}{N!} & \text{if } \rho = \rho_\mr{d} ,
 \end{cases}
\end{align}
where we have used the fact that any projector onto irreps is unitarily invariant.
These then are universal upper bounds on the success probability for singly and completely distinguishable states, respectively.
However, since we are restricted to photon number counting measurements, we will see that while the first bound is achievable, the second is not in general.
We will go through various examples in detail in the following sections.

\subsection{Two modes}
\label{subsec:two}

\subsubsection{Two photons in two modes}
\label{subsec:twotwo}

In the case of two photons in two modes, the states to be discriminated are, from Eqs.~(\ref{eq:rhoi}), (\ref{eq:rhos}) and (\ref{eq:rhod}),
\begin{align}
\rho_{\mr{i}} &=  \Ket{\scriptsize\young(12)\,} \Bra{\scriptsize\young(12)\,}, \quad \mr{and} \\
\rho_{\mr{s}} &= \rho_{\mr{d}}  = \frac{1}{2}\Ket{\scriptsize\young(12)\,} \Bra{\scriptsize\young(12)\,} + \frac{1}{2}\Ket{\scriptsize\young(1,2)\,} \Bra{\scriptsize\young(1,2)\,}.
\end{align}
Observing that there is only one available state which is not symmetric, it is easy to write down an arbitrary \emph{partially} distinguishable System state in this case, since there is but one parameter:
\begin{align}
\rho &= \alpha \Ket{\scriptsize\young(12)\,} \Bra{\scriptsize\young(12)\,} + (1-\alpha)\Ket{\scriptsize\young(1,2)\,} \Bra{\scriptsize\young(1,2)\,} .
\end{align}
As discussed in Sec.~\ref{subsec:simp}, only occupations that do not have all the photons bunched in the same mode can be used for meaningful discrimination, in this case leaving only one choice of projector, the coincidence $M_{(1,1)} = \Ket{\scriptsize\young(12)\,} \Bra{\scriptsize\young(12)\,} + \Ket{\scriptsize\young(1,2)\,} \Bra{\scriptsize\young(1,2)\,}$.

In our discussion in Sec.~\ref{sec:background}, we claimed that the optimal discriminator is given by a coincidence count and a balanced beamsplitter; we can now prove this assertion.
First, note that since there is only one antisymmetric state, the antisymmetric irreducible representation of any $U$ has but one matrix element and so the action of any interferometer  on this state is trivial (in Eq.~(\ref{eq:perdet}) given by its determinant).
Thus the only contribution to non-symmetric part of Eq.~(\ref{eq:simplifiedFD}) is $\abs{\Bra{\scriptsize\young(1,2)\,} U^{\yoneone} \Ket{\scriptsize\young(1,2)\,}} = 1$, and there is nothing to maximise in Eq.~(\ref{eq:Allmax}).
All that is left is to satisfy the constraint, Eq.~(\ref{eq:Allcnstrt}).
Parametrising $U$ as
\begin{equation}
\begin{bmatrix}
e^{\iu \phi} \cos{\theta} & e^{\iu \varphi}  \sin{\theta} \\
-e^{-\iu \varphi} \sin{\theta} &  e^{-\iu \phi} \cos{\theta}
\end{bmatrix}.
\label{eq:Uparam22}
\end{equation}
one finds that the constraint is then per$U = \cos^2{\theta} - \sin^2{\theta} = \cos{2 \theta} = 0$,
with the family of solutions $\{ (\phi, \varphi, \pi/4) | \, 0 \leq \phi \leq \pi, 0 \leq \varphi \leq \pi \}$.
The solutions do not depend on the phases $\phi$ or $\varphi$, as we would expect from the discussion in Sec.~\ref{subsec:Reck}, but only on the choice of the beamsplitter reflectivity, which is balanced as claimed.

We see that not only does unambiguous discrimination return the HOM measurement as was discussed in Sec.~\ref{subsec:HOM}, it is optimal for an arbitrary partially distinguishable two photon state.

\subsubsection{Three photons in two modes}

As an example of the utility of the formalism, in this subsection we consider the simplest nontrivial case with $N(=3)>d_\mr{S}(=2)$.
As mentioned in Sec.~\ref{sec:states}, this restricts the kinds of distinguishable states that can occur; we consider situations with two photons in one System mode and the third in the other.
The indistinguishable state is $\crop{a}{11}\crop{a}{11}\crop{a}{21}\ket{\text{vac}}={\tiny \Ket{\setlength\arraycolsep{2pt}\renewcommand{\arraystretch}{0.75}\begin{matrix} 2\\1 \end{matrix}}}$, with reduced state
\begin{equation}
\rho_{\mr{i}} = \Ket{\scriptsize\young(112)\,} \Bra{\scriptsize\young(112)\,}.
\end{equation}
There are essentially two types of distinguishable state in this situation.
The first is $\crop{a}{11}\crop{a}{11}\crop{a}{22} \ket{\text{vac}} ={\tiny \Ket{\setlength\arraycolsep{2pt}\renewcommand{\arraystretch}{0.75}\begin{matrix} 2&0\\0&1 \end{matrix}}}$, and the second $\crop{a}{11}\crop{a}{12}\crop{a}{21} \ket{\text{vac}}={\tiny \Ket{\setlength\arraycolsep{2pt}\renewcommand{\arraystretch}{0.75}\begin{matrix} 1&1\\1&0 \end{matrix}}}$.
Other states are equivalent to the above for the reasons discussed in Sec.~\ref{subsec:singlyD}.
Further, the (now incompletely) distinguishable state $\crop{a}{11}\crop{a}{12}\crop{a}{23}\ket{\text{vac}}={\tiny \Ket{\setlength\arraycolsep{2pt}\renewcommand{\arraystretch}{0.75}\begin{matrix} 1&1&0\\0&0&1 \end{matrix}}}$ has a reduced state that is the same as Eq.~(\ref{eq:rhod1}), and will therefore have the same discrimination measurement and success probability.
The reduced state for the first case is
\begin{align}
\rho_{\mr{s}_1} &= \frac{1}{3} \Ket{\scriptsize\young(112)\,} \Bra{\scriptsize\young(112)\,}\nonumber\\
&\quad+ \frac{1}{3} \Ket{\scriptsize\young(11,2)_{1}} \Bra{\scriptsize\young(11,2)_{1}}
+ \frac{1}{3} \Ket{\scriptsize\young(11,2)_{2}} \Bra{\scriptsize\young(11,2)_{2}}  , \label{eq:rhod1}
\end{align}
while that for the second case is
\begin{align}
\rho_{\mr{s}_2} &= \frac{4}{6} \Ket{\scriptsize\young(112)\,} \Bra{\scriptsize\young(112)\,}\nonumber\\
&\quad+ \frac{1}{6} \Ket{\scriptsize\young(11,2)_{1}} \Bra{\scriptsize\young(11,2)_{1}} 
+ \frac{1}{6} \Ket{\scriptsize\young(11,2)_{2}} \Bra{\scriptsize\young(11,2)_{2}}  . \label{eq:rhod2}
\end{align}
Note that Eq.~(\ref{eq:Allmax}) does not depend on the amplitude of the symmetric part of the state -- its contribution has to be zero by  Eq.~(\ref{eq:Allcnstrt}).
It only depends on the nonsymmetric components, and since $\rho_{\mr{s}_1}$ and $\rho_{\mr{s}_2}$ are equally weighted across the available nonsymmetric states, the optimal discriminator will be the same.
However $\rho_{\mr{s}_2}$ does have half of the amplitude of  $\rho_{\mr{s}_1}$ in this subspace, which will halve the success probability.

There are four possible occupations to measure, however as mentioned in Sec.~\ref{subsec:simp} the bunched ones can be disregarded and the optimisation carried out on $M_{(2,1)}$ and $M_{(1,2)}$.
We parametrise $U$ again as in Eq.~(\ref{eq:Uparam22}).
For  $M_{(2,1)}$ Eq.~(\ref{eq:Allcnstrt}) reduces to $\abs{\Bra{\scriptsize\young(112)\,} U^{\ythree} \Ket{\scriptsize\young(112)\,}} = \abs{ (\cos{\theta} + 3 \cos{3\theta})/4} = 0$.
Since $0 \leq \theta \leq \pi$, this equation is true for $\theta \in \{ \pi/2, \,\arccos{(\sqrt{2/3})}, \,\arccos{(-\sqrt{2/3})} \}$.
On the other hand, Eq.~(\ref{eq:Allcnstrt}) for $M_{(1,2)}$ is $\abs{\Bra{\scriptsize\young(122)\,} U^{\ythree}  \Ket{\scriptsize\young(112)\,}} =  \abs{ (\sin{\theta} - 3 \cos{3\theta})/4}$.
This equation cannot be zero for the above choice of angles that ensure $\abs{\Bra{\scriptsize\young(112)\,} U^{\ythree}  \Ket{\scriptsize\young(112)\,}} = 0 $.
Thus, only one of the outcomes can be used to discriminate these states; without loss of generality, we choose to optimise for $M_{(2,1)}$.
In this case we want to maximise $\Tr \left[ \rho_{\mr{s}_1} M_{(2,1)} (U) \right] = 2 \abs{\Bra{\scriptsize\young(11,2)_{1}} U^{\ytwoone} \Ket{\scriptsize\young(11,2)_{1}} }^2 =  2 \cos^2{\theta} /3$.
When $\theta = \pi / 2$, we get success probability of $0$.
When $\theta = \pm \arccos{(\sqrt{2/3})}$, we get success probability of $4/9$.
Thus an optimal discriminating interferometer is $U={\tiny \begin{bmatrix} \sqrt{2} & 1 \\ -1 & \sqrt{2} \end{bmatrix}/\sqrt{3}}$, with success probabilities $4/9$ for $\rho_{\mr{s}_1}$ and $2/9$ for $\rho_{\mr{s}_2}$.

\subsection{Three modes}
\label{subsec:three}

From now on we will only consider coincident input with $N=d_\mr{S}$.
For three photons in three System modes, the completely indistinguishable reduced state, is from Eq.~(\ref{eq:rhoi}), 
\begin{equation}
\rho_{\mr{i}} = \ket{(3),\underline{1}} \bra{(3),\underline{1}} = \Ket{\scriptsize\young(123)\,}\Bra{\scriptsize\young(123)\,} . \label{eq:rhoi3}
\end{equation}

There are now three different singly distinguishable states, depending on which System mode the `bad' photon is in. 
In the Schur-Weyl basis (see Sec.~\ref{subsec:SchurWeylImplement}) their full System-Label states, as per the discussion in Sec.~\ref{subsec:singlyD}, are
\begin{align}
\sqrt{3} \, \crop{a}{11}\crop{a}{21}\crop{a}{32} \ket{\text{vac}} 
=& \sqrt{3} {\tiny \Ket{\setlength\arraycolsep{2pt}\renewcommand{\arraystretch}{0.75}\begin{matrix} 1&0\\1&0\\0&1 \end{matrix}}} \nonumber\\
=&  \Ket{\scriptsize\young(123)\,} \Ket{\scriptsize\young(112)\,} \nonumber\\
&+  \Ket{\scriptsize\young(12,3)_1} \Ket{\scriptsize\young(11,2)_1} \nonumber\\
&+  \Ket{\scriptsize\young(12,3)_2} \Ket{\scriptsize\young(11,2)_2}, \label{eq:singleThreeOne} \\
\sqrt{3} \, \crop{a}{11}\crop{a}{22}\crop{a}{31} \ket{\text{vac}}
=& \sqrt{3} {\tiny \Ket{\setlength\arraycolsep{2pt}\renewcommand{\arraystretch}{0.75}\begin{matrix} 1&0\\0&1\\1&0 \end{matrix}}} \nonumber\\
=&  \Ket{\scriptsize\young(123)\,} \Ket{\scriptsize\young(112)\,} \nonumber\\
&- \frac{1}{2} \left(  \Ket{\scriptsize\young(12,3)_1}  +  \sqrt{3}   \Ket{\scriptsize\young(13,2)_1} \right) \Ket{\scriptsize\young(11,2)_1} \nonumber\\
&- \frac{1}{2} \left(  \Ket{\scriptsize\young(12,3)_2} + \sqrt{3}    \Ket{\scriptsize\young(13,2)_2} \right) \Ket{\scriptsize\young(11,2)_2}  , \label{eq:singleThreeTwo} \\
\sqrt{3} \, \crop{a}{12}\crop{a}{21}\crop{a}{31} \ket{\text{vac}} 
=& \sqrt{3} {\tiny \Ket{\setlength\arraycolsep{2pt}\renewcommand{\arraystretch}{0.75}\begin{matrix} 0&1\\1&0\\1&0 \end{matrix}}} \nonumber\\
=&  \Ket{\scriptsize\young(123)\,} \Ket{\scriptsize\young(112)\,}  \nonumber\\
&- \frac{1}{2}  \left(  \Ket{\scriptsize\young(12,3)_1}  -  \sqrt{3}   \Ket{\scriptsize\young(13,2)_1} \right) \Ket{\scriptsize\young(11,2)_1} \nonumber\\
&- \frac{1}{2} \left( \Ket{\scriptsize\young(12,3)_2} - \sqrt{3}   \Ket{\scriptsize\young(13,2)_2} \right) \Ket{\scriptsize\young(11,2)_2} . \label{eq:singleThreeThree}
\end{align}
While for completely distinguishable states permuting System modes has no effect on the reduced state, here the reduced states will not be invariant.
However, because permutations of System modes lie inside the set of allowed operations, (that is, S$_{d_\mr{S}} \subset \mr{U}(d_\mr{S})$), if we optimise for one of these states, the resulting interferometer will be easily related to the others by including some mode swapping.
Therefore we can focus on one of these states and the success probabilities that we find will be the same for the other two; Eq.~(\ref{eq:singleThreeOne}) has the reduced state (cf. Eq.~(\ref{eq:rhos}))
\begin{align}
\rho_{\mr{s}} =& \frac{1}{3} \Ket{\scriptsize\young(123)\,}\Bra{\scriptsize\young(123)\,} \nonumber \\
&+ \frac{1}{3} \Ket{\scriptsize\young(12,3)_1\,}\Bra{\scriptsize\young(12,3)_1\,} + \frac{1}{3} \Ket{\scriptsize\young(12,3)_2\,}\Bra{\scriptsize\young(12,3)_2\,} . \label{eq:rhos3}
\end{align}
It is natural to ask about discrimination of mixtures of these three states; we will discuss this in Sec.~\ref{subsec:mixedthree}.

The completely distinguishable state corresponding to $\crop{a}{11}\crop{a}{22}\crop{a}{33}\ket{\text{vac}}={\tiny \Ket{\setlength\arraycolsep{2pt}\renewcommand{\arraystretch}{0.75}\begin{matrix} 1&0&0\\ 0&1&0 \\ 0&0&1 \end{matrix}}}$ per Eq.~(\ref{eq:rhod}) is
\begin{align}
\rho_{\mr{d}} =& \frac{1}{6} \Ket{\scriptsize\young(123)\,}\Bra{\scriptsize\young(123)\,} + \frac{1}{6} \Ket{\scriptsize\young(1,2,3)\,}\Bra{\scriptsize\young(1,2,3)\,} \nonumber \\
&+ \frac{1}{6} \Ket{\scriptsize\young(12,3)_1\,}\Bra{\scriptsize\young(12,3)_1\,} + \frac{1}{6} \Ket{\scriptsize\young(12,3)_2\,}\Bra{\scriptsize\young(12,3)_2\,} \nonumber \\
&+ \frac{1}{6} \Ket{\scriptsize\young(13,2)_1\,}\Bra{\scriptsize\young(13,2)_1\,} + \frac{1}{6} \Ket{\scriptsize\young(13,2)_2\,}\Bra{\scriptsize\young(13,2)_2\,} . \label{eq:rhod3}
\end{align}

For the following let us define two sets of measurement operators: those with two photons in one mode, $M_{(2,1,0)} = \sum_{\lambda\neq\yoneoneone,p} \ket{\lambda,p,(2,1,0)} \bra{\lambda,p,(2,1,0)}$, $M_{(2,0,1)}$, $M_{(1,0,2)}$, $M_{(1,2,0)}$, $M_{(0,1,2)}$, and $M_{(0,2,1)}$, which we denote $\mathcal{M}_2$; and those with each photon in a different mode, that is $\mathcal{M}_1 \ni M_{(1,1,1)} = \sum_{\lambda,p,r} \ket{\lambda,p,\underline{1},r} \bra{\lambda,p,\underline{1},r}$.
As discussed in Sec.~\ref{subsec:simp}, the measurements $M_{(3,0,0)} = \Ket{\scriptsize\young(111)\,}\Bra{\scriptsize\young(111)\,}$, $M_{(0,3,0)}$, and $M_{(0,0,3)}$ will not be helpful for discrimination.

\subsubsection{Discriminating singly distinguishable states}

\begin{figure}[h]
\includegraphics[width=0.45\textwidth]{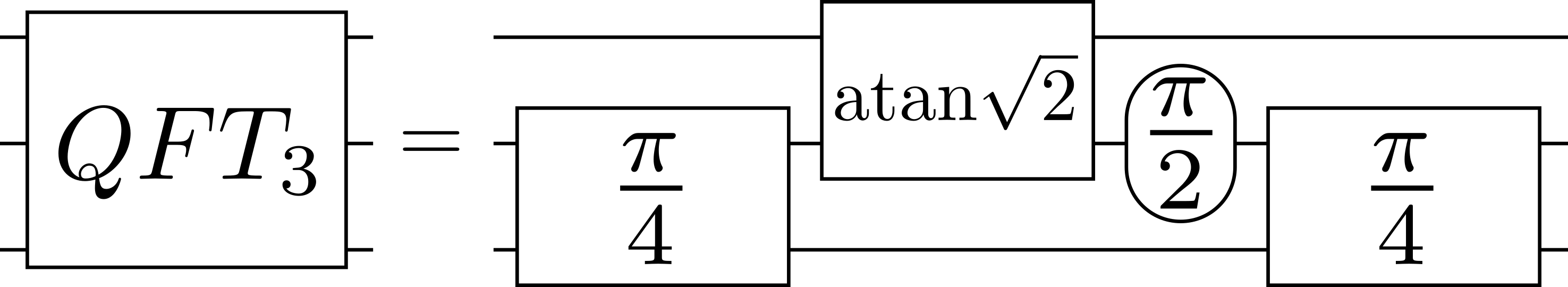}
\caption{
The best known interferometer for discriminating completely indistinguishable from distinguishable states of three photons in three modes is $QFT_3$, with a success probability of $2/3$.
Up to phases, it consists of two balanced beamsplitters, one $2:1$ beamsplitter, and one $\pi/2$ phaseshifter.
}
\label{fig:opt33}
\end{figure}

Let $\rho^{\lambda}$ denote the (unnormalized) part of a state supported on the subspace of irrep $\lambda$.
Notice that $\rho_{\mr{s}}$ has no support in the antisymmetric subspace, so that $\sum_{\underline{n}} \Tr{\left[  \rho_{\mr{s}}^{\ytwoone} M_{\underline{n}} (U)  \right]} = 2 / 3$ and $\sum_{\underline{n}} \Tr{\left[  \rho_{\mr{s}}^{\yoneoneone} M_{\underline{n}} (U)  \right]} = 0$.
In the best case scenario, we can pick some subset of occupations $D$ and $U$ for which Eq.~(\ref{eq:Allcnstrt}) holds and the success probability will be bounded by $2/3$.
It is well known how to achieve this; use a balanced tritter, $U = QFT_3$, and all the occupations from $\mathcal{M}_2$, where $QFT_N$ is defined as
\begin{equation}
QFT_N = \frac{1}{\sqrt{N}} \begin{bmatrix}
1 & 1 & \cdots & 1 \\
1 & \omega^1  & \cdots &  \omega^{N-1} \\
\vdots & \vdots & & \vdots \\
1 &   \omega^{N-1} &  \cdots & \omega^{(N-1)(N-1)}
\end{bmatrix}
\end{equation}
and $\omega = \exp{{\frac{2 \pi \iu}{N}}}$.
A parametrisation that realizes a balanced tritter is given in Figure~\ref{fig:opt33}.

\subsubsection{Discriminating mixed singly distinguishable states}
\label{subsec:mixedthree}

A short digression regarding mixed System-Label states: if we were (uniformly) ignorant about which mode the `bad' photon was in, we would have an equal mixture of Eqs.~(\ref{eq:singleThreeOne}, \ref{eq:singleThreeTwo}, \ref{eq:singleThreeThree}).
The resulting mixed state is
\begin{align}
\rho_{\mr{sm}}
:=& \frac{1}{3} \Ket{\scriptsize\young(123)\,} \Bra{\scriptsize\young(123)\,}    \nonumber \\
&+ \frac{1}{6} \Ket{\scriptsize\young(12,3)_1\,} \Bra{\scriptsize\young(12,3)_1\,} + \frac{1}{6} \Ket{\scriptsize\young(12,3)_2\,} \Bra{\scriptsize\young(12,3)_2\,} \nonumber \\
&+ \frac{1}{6} \Ket{\scriptsize\young(13,2)_1\,} \Bra{\scriptsize\young(13,2)_1\,} + \frac{1}{6} \Ket{\scriptsize\young(13,2)_2\,} \Bra{\scriptsize\young(13,2)_2\,} .
\label{eq:singleThreeMixed}
\end{align}
The overlap $\sum_{\underline{n}} \Tr{\left[  \rho_{\mr{sm}}^{\ytwoone} M_{\underline{n}} (U)  \right]} = 2 / 3$ for $\mathcal{M}_2$ again saturates the bound, and a balanced tritter remains the best choice of interferometer.
This can be seen from the symmetry of the $QFT$ which treats a `bad' photon in any mode essentially the same way, and so should be true for analogous singly distinguishable mixed states for all $N$, however we will not discuss mixed System-Label states further here.

\subsubsection{Discriminating completely distinguishable states}

Using the cost function from Eq.~(\ref{eq:costfnFD}) and a range of penalties $\xi \in \{2,4,6,8,10\}$ we found that the highest success probability in the completely distinguishable case is $2/3$.
The measurement operators were always the full set $\mathcal{M}_2$ with a balanced tritter as an example solution, just as in the previous section.
However, this does not saturate the bound in Sec.~\ref{subsec:bounds}, which is $5/6$ in the case of three photons.
To investigate this further, we try to understand the structure of the state a bit better and use numerical evidence to show that a balanced tritter is likely to be optimal.

From Eq.~(\ref{eq:rhod3}) we have $\sum_{\underline{n}} \Tr{\left( \rho_{\mr{d}}^{\ytwoone}  M_{\underline{n}} (U)  \right)} = 2 / 3$ and $\sum_{\underline{n}} \Tr{\left( \rho_{\mr{d}}^{\yoneoneone} M_{\underline{n}} (U)  \right)} = 1 / 6$, so that $\sum_{\underline{n}} \Tr{\left( \rho_{\mr{d}}^{\overline{\ythree}} M_{\underline{n}} (U) \right)} = 5 / 6$, which is the discrimination bound.
Notice that operators from $\mathcal{M}_2$ do not have support on the anti-symmetric subspace.
Therefore, if we only pick operators from  $\mathcal{M}_2$ as the discriminating operators, and assume they can simultaneously satisfy Eq.~(\ref{eq:Allcnstrt}), then  $\sum_{\underline{n} \in \mathcal{M}_2} \Tr{\left(  \rho_{\mr{d}} M_{\underline{n}} (U) \right)} = \sum_{\underline{n} \in \mathcal{M}_2} \Tr{\left(\rho_{\mr{d}}^{\ytwoone} M_{\underline{n}} (U) \right)} \leq 2 / 3$.
This is exactly what happens for the interferometers from our optimisation.

This tells us that if we want the success probability to be larger than $2/3$, the only operator left, $M_{(1,1,1)}$, would have to be included.
Our numerical results show that, on the contrary, it is unlikely for any $D$ that includes $M_{(1,1,1)}$ to give a success probability over $1/2$.
We do this with a new cost function, much like Eq.~(\ref{eq:costfnFD}) but modified to force $M_{(1,1,1)}$ to be included:
\begin{align}
C_{\mr{d},111}(U) = \eta \Tr(\rho_{\mr{i}} M_{(1,1,1)}(U)) + 
C_{\mr{d}}(U) , \label{eq:threephotoncostfn1}
\end{align}
where $\eta $ is a penalty to ensure Eq.~(\ref{eq:Allcnstrt}) for $M_{(1,1,1)}$ has to be satisfied, and $C_{\mr{d}}(U)$ is as defined in Eq.~(\ref{eq:costfn}).
This penalty is set to $\eta = 10$ making the first term an order of magnitude higher than the second term of Eq.~(\ref{eq:threephotoncostfn1}), where we took $\xi = 6$.
As we learned in Sec.~\ref{subsec:Reck}, we can ignore the outside phaseshifters of the standard Reck parametrisation, therefore we are only optimizing over $4$ parameters, $\theta_{2,1},\theta_{2,2},\theta_{1,2}$, and $\omega_{1,2}$.
The lowest value of the cost function found by the optimisation techniques in Sec.~\ref{subsec:numerical} is $-0.500426$.
This corresponds to a success probability of $0.5$ in discriminating between the two states, which is lower than the $2/3$ achievable when $M_{(1,1,1)}$ is not included.

While this does not give us definitive proof that no scheme that includes a threefold coincidence can give success probability higher than $2/3$, it does strongly indicate that this should be true.
Moreover, with the same optimisation functions we investigated how many of the other operators alongside $M_{(1,1,1)}$ we can pick at the same time, and it seems that the best we can do is to have four from $\mathcal{M}_2$ satisfy Eq.~(\ref{eq:Allcnstrt}) simultaneously.
However, in all the situations when this occurs, some of the terms in Eq.~(\ref{eq:Allmax}) are zero, thus the success probability remains at $1/2$, which can be achieved using just $M_{(1,1,1)}$ and a balanced beamsplitter.

The balanced tritter uses all the measurement operators from $\mathcal{M}_2$, with each contributing $1/9$ to achieve the success probability $2/3$.
To draw attention to the difference between optimizing a single operator and multiple operators at once, mentioned in Sec.~\ref{subsec:discrimination}, we notice that optimizing for one operator from the set $\mathcal{M}_2$ yields a success probability higher than $1/9$ (for some other choice of $U$).
Taking this further, we can search numerically for the single best outcome, with a cost function similar to that of Eq.~(\ref{eq:costfnFD}), except we now focus only on a single $\underline{n}$, that is $C(\underline{n}, U) = - 2 \sum_{\lambda, r,r'} \allowbreak \exp \left(- \xi |\braket{(3),\underline{n}|U|(3),\underline{1}}|^{2} \right)  |\braket{\lambda,\underline{n},r'|U|\lambda,\underline{1},r}|^2$.
We find $M_{(1,1,1)}$ is a clear winner with a total success probability of $1/2$, achievable by a balanced beamsplitter as mentioned above.
All of the other operators by themselves only ever give an optimised success probability of $1/8$.
Notice that $6 \cdot{} 1/8 = 3/4 > 2 / 3$, showing that the strategy that gives us the best success chance with a single operator from $\mathcal{M}_2$ can not be achieved simultaneously by all six of them.

\subsection{Four and more modes}

\setlength{\fboxrule}{0pt}
\begin{table*}[p]
\centering
\begin{tabular}{ | >{\centering\arraybackslash}m{0.5cm} || >{\centering\arraybackslash}m{4.5cm} || >{\centering\arraybackslash}m{2cm} |  >{\centering\arraybackslash}m{1cm} | >{\centering\arraybackslash}m{1cm} | >{\centering\arraybackslash}m{1cm} || >{\centering\arraybackslash}m{2.2cm} | >{\centering\arraybackslash}m{2.2cm}|}
\hline
 &  & \multicolumn{4}{c||}{Singly distinguishable, $\rho_{\mr{s}}$ } & \multicolumn{2}{c|}{Completely distinguishable, $\rho_{\mr{d}}$ }  \\
\cline{3-8}
$N$  & $U$ &  \multicolumn{4}{c||}{Success probability} & \multicolumn{2}{c|}{Success probability} \\ 
 &  & Bound & Best & Worst & Avg & Bound & \\ \hline \hline
2   & \fbox{\includegraphics[width=0.10\textwidth]{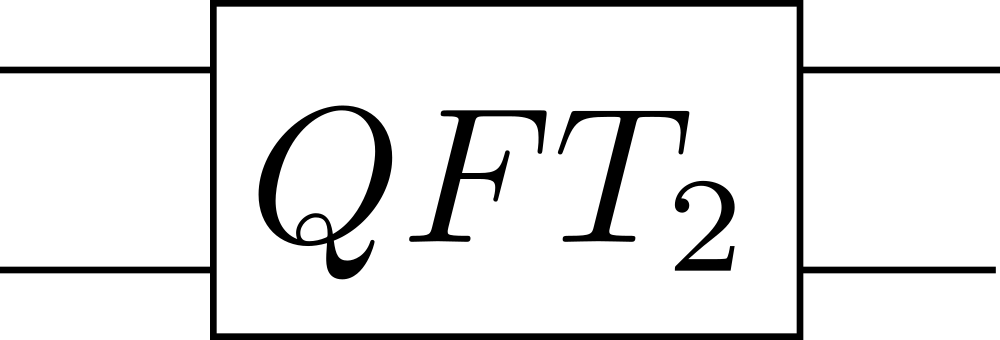}}  & $\frac{1}{2}=0.5000$   & \multicolumn{3}{c||}{$\frac{1}{2}=0.5000$}& $\frac{1}{2} = 0.5000$ & $\frac{1}{2} = 0.5000$ \\ \hline
3   & \fbox{\includegraphics[width=0.10\textwidth]{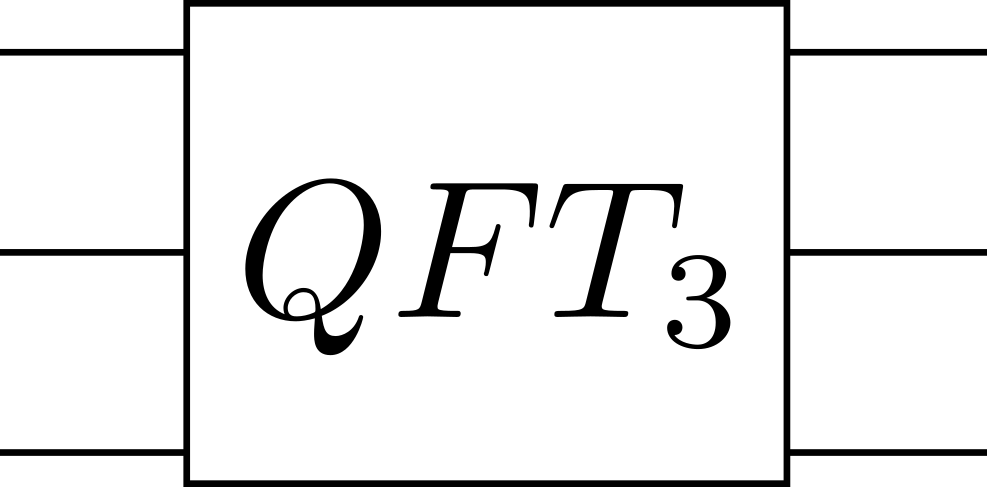}}  & $\frac{2}{3} \approx 0.6666$   & \multicolumn{3}{c||}{$\frac{2}{3} \approx 0.6666$} & $\frac{5}{6} \approx 0.8333$ &  $\frac{2}{3} \approx  0.6666$* \\ \hline
\multirow{2}{*}[-1em]{4}   & \fbox{\includegraphics[width=0.17\textwidth]{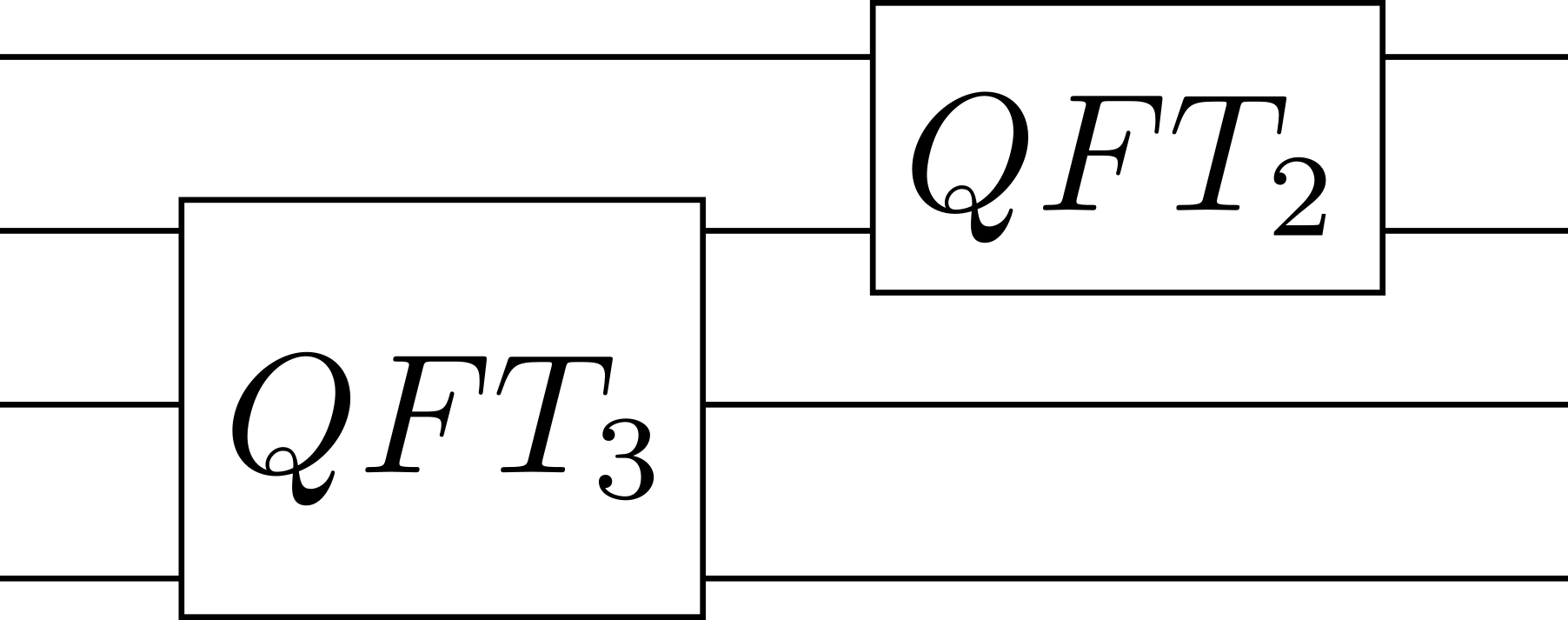}}  & \multirow{2}{*}[-1em]{$\frac{3}{4}=0.7500$} & $\frac{25}{36} \approx 0.6944$ & $\frac{1}{4}=0.2500$  & $\frac{7}{12} \approx 0.5833$ & \multirow{2}{*}[-1em]{$\frac{23}{24} \approx 0.9583$} &  $\frac{19}{24} \approx 0.7916$* \\ \cline{2-2}\cline{4-6}\cline{8-8} 
                     & \fbox{\includegraphics[width=0.10\textwidth]{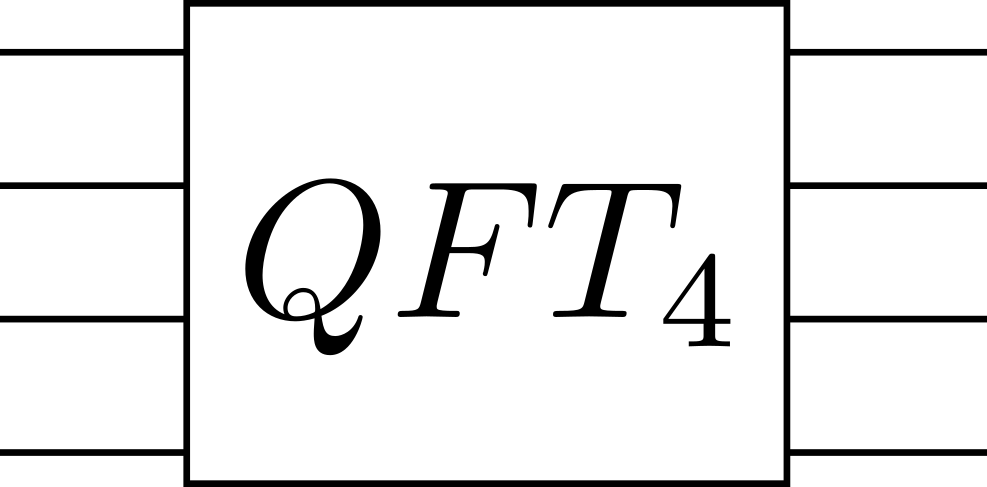}} &                     & \multicolumn{3}{c||}{$\frac{3}{4}=0.7500$} &   & $\frac{3}{4} = 0.7500$ \\ \hline
\multirow{2}{*}[-1em]{5}   & \fbox{\includegraphics[width=0.17\textwidth]{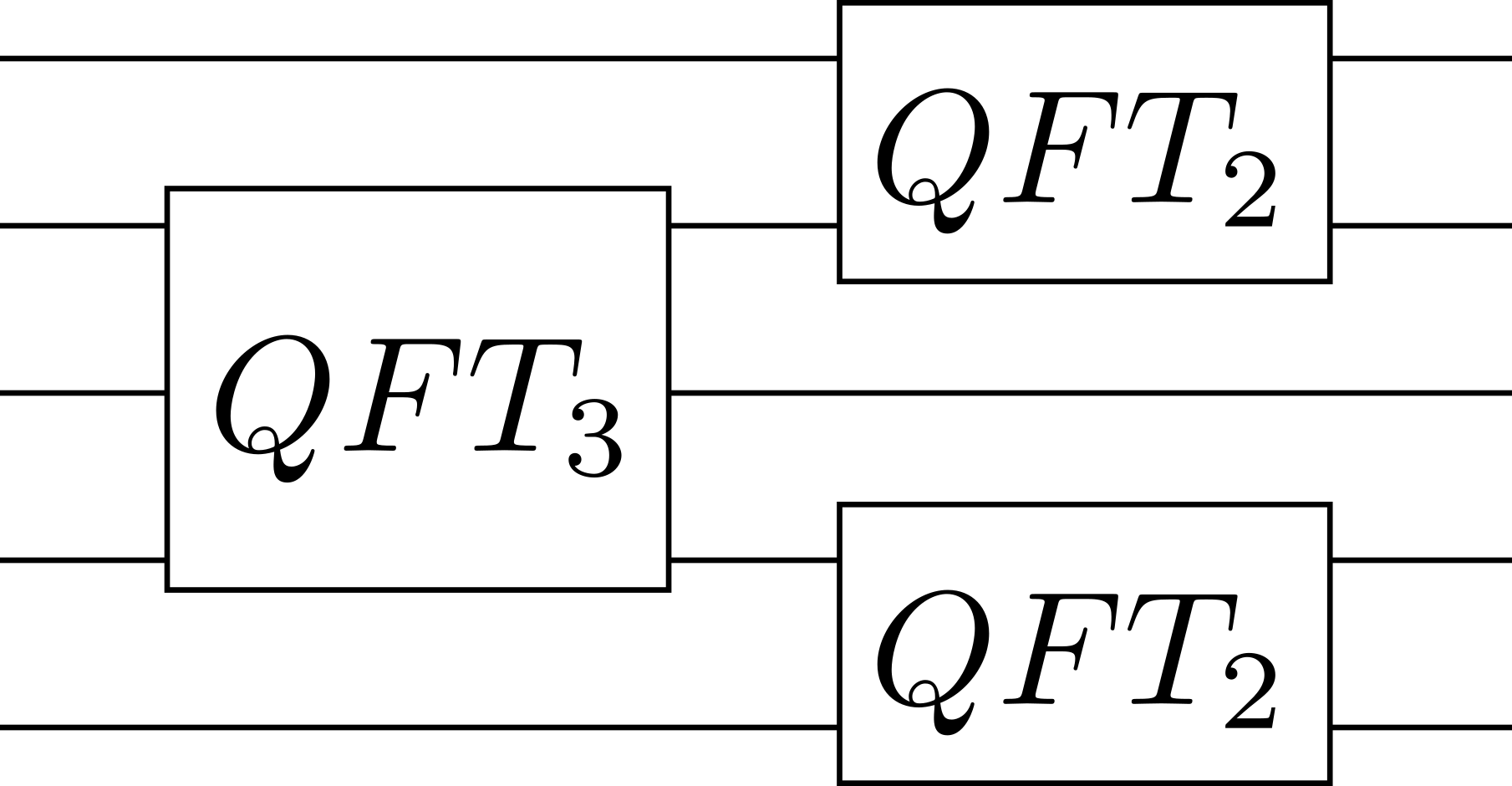}}  & \multirow{2}{*}[-1em]{$\frac{4}{5}=0.8000$}  & $\frac{8}{15} \approx 0.7222$ & $\frac{1}{4}=0.2500$ & $\frac{13}{18} \approx 0.5333$ & \multirow{2}{*}[-1em]{$\frac{119}{120} \approx 0.9917$} & $\frac{31}{36} \approx 0.8611$* \\ \cline{2-2}\cline{4-6}\cline{8-8} 
                     & \fbox{\includegraphics[width=0.10\textwidth]{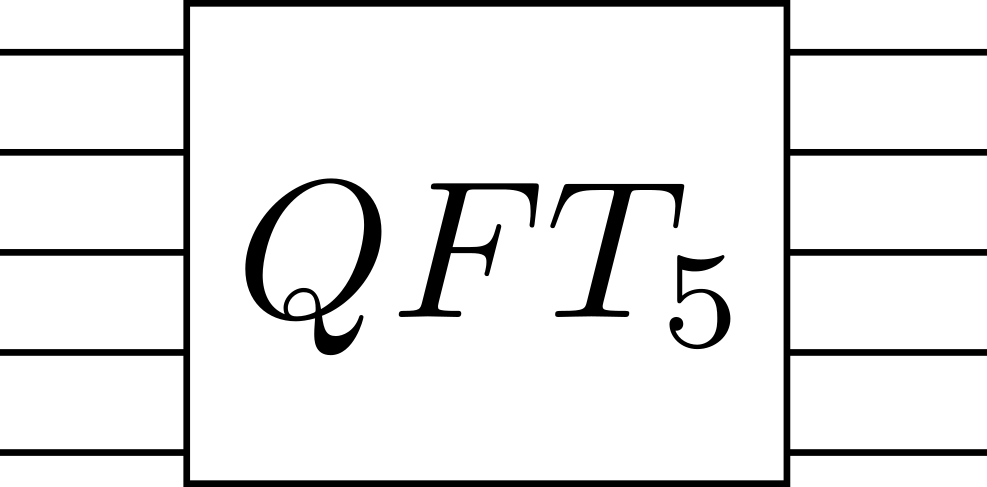}} &                     & \multicolumn{3}{c||}{$\frac{4}{5}=0.8000$} &   & $\frac{4}{5} = 0.8000$ \\ \hline
\multirow{2}{*}[-1em]{6}   & \fbox{\includegraphics[width=0.17\textwidth]{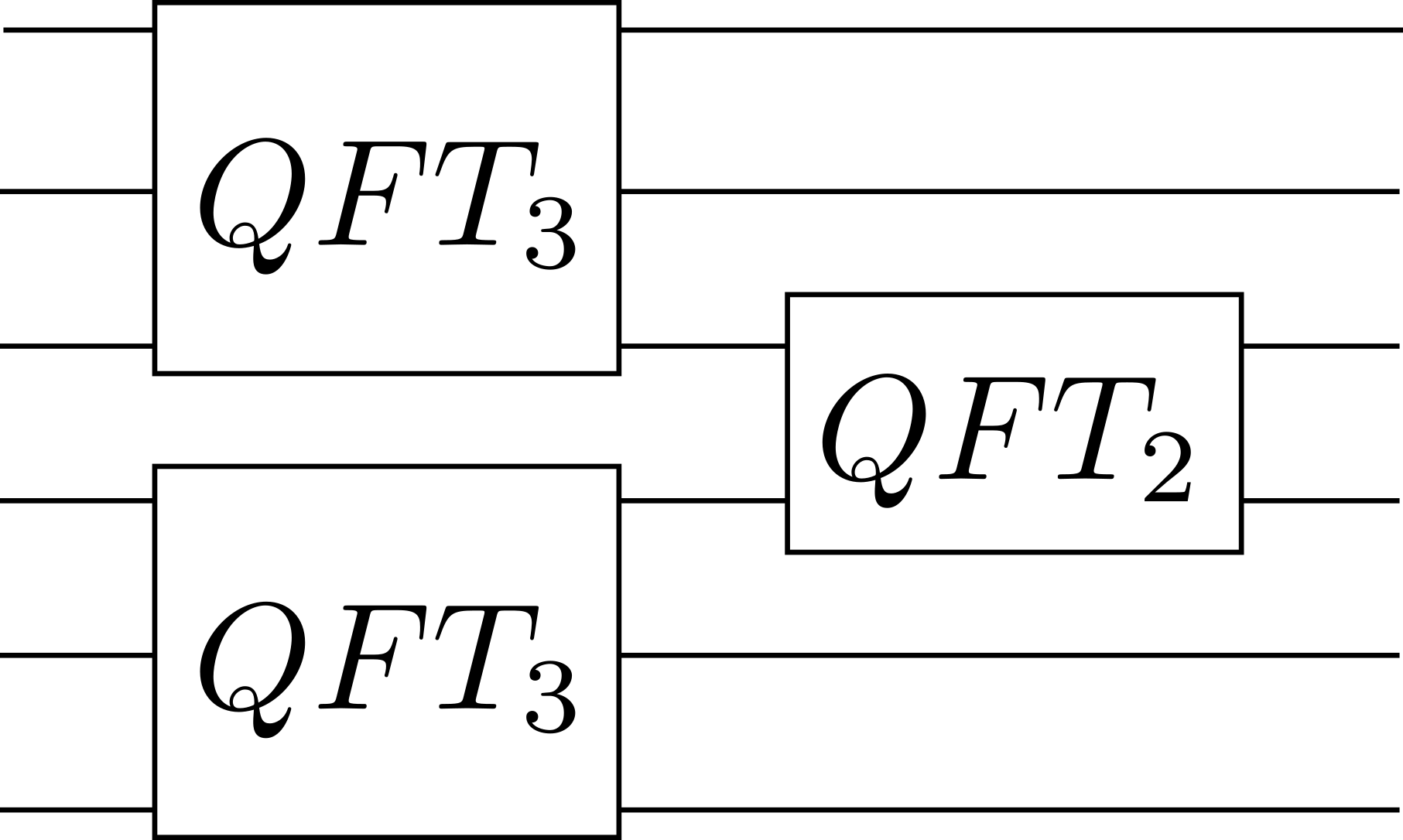}}  & \multirow{2}{*}[-1em]{$\frac{5}{6} \approx 0.8333$}  & $\frac{167}{243} \approx 0.6872$ & $\frac{167}{243} \approx 0.6872$  & $ \frac{167}{243} \approx 0.6872$ & \multirow{2}{*}[-1em]{$\frac{719}{720} \approx 0.9986$} & $\frac{671}{729} \approx 0.9204$ \\ \cline{2-2}\cline{4-6}\cline{8-8} 
                     & \fbox{\includegraphics[width=0.10\textwidth]{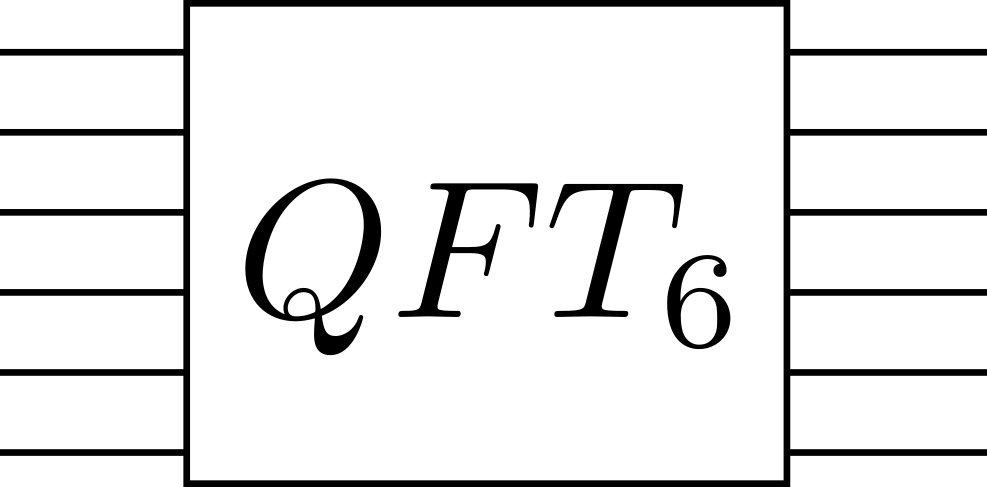}} &                     & \multicolumn{3}{c||}{$\frac{5}{6} \approx 0.8333$} &   & $\frac{65}{72} \approx 0.9028$ \\ \hline
\multirow{2}{*}[-1em]{7}   & \fbox{\includegraphics[width=0.17\textwidth]{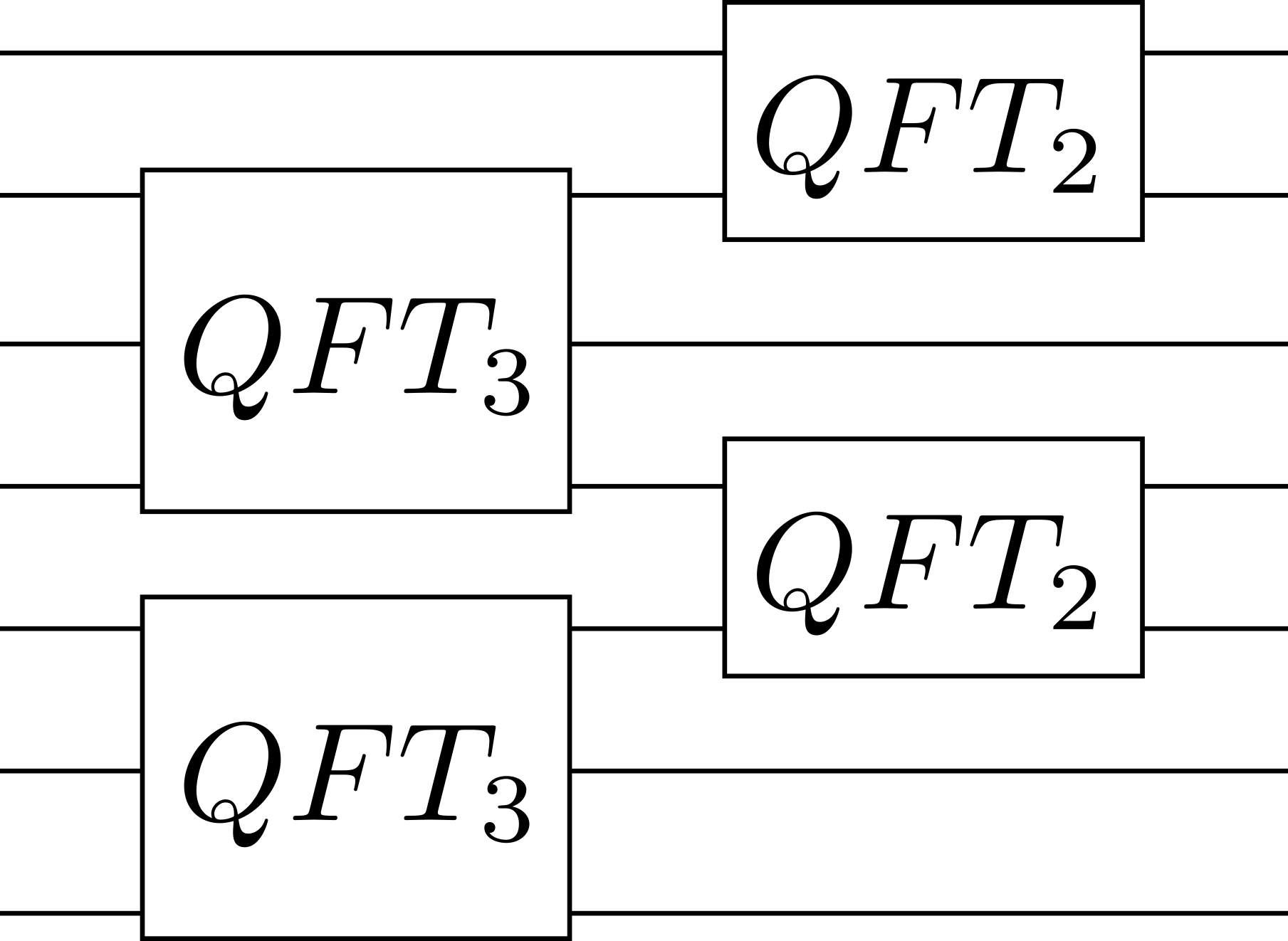}}  & \multirow{2}{*}[-1em]{$\frac{6}{7} \approx 0.8571$}   &  $\frac{695}{972} \approx 0.7150$ &   $\frac{1}{4} = 0.2500$  & $\frac{361}{567} \approx 0.6367$ & \multirow{2}{*}[-1em]{$\frac{5039}{5040} \approx 0.9998$}  & $\frac{2765}{2916} \approx 0.9482$ \\ \cline{2-2}\cline{4-6}\cline{8-8} 
                     & \fbox{\includegraphics[width=0.10\textwidth]{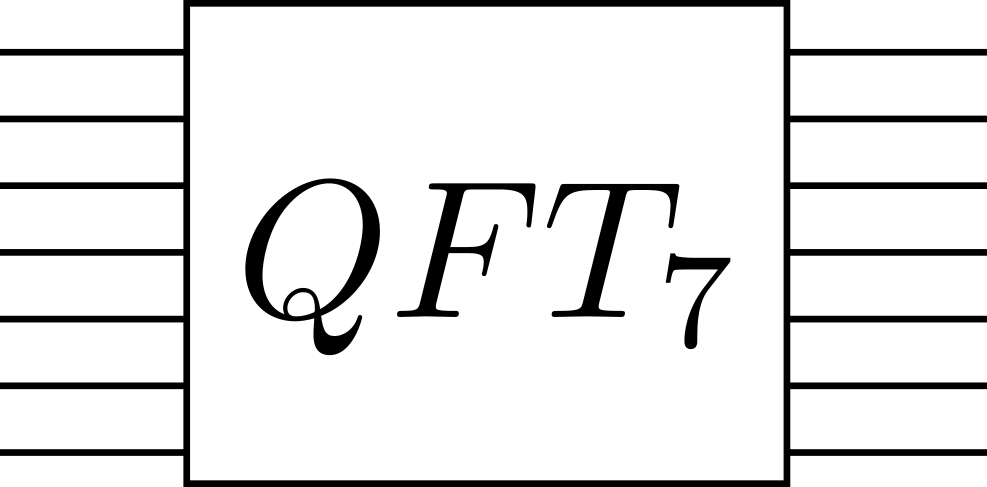}} &                     & \multicolumn{3}{c||}{$\frac{6}{7} \approx 0.8571$} &   & $\frac{6}{7} \approx 0.8571$\\ \hline
\multirow{2}{*}[-1em]{8}   & \fbox{\includegraphics[width=0.17\textwidth]{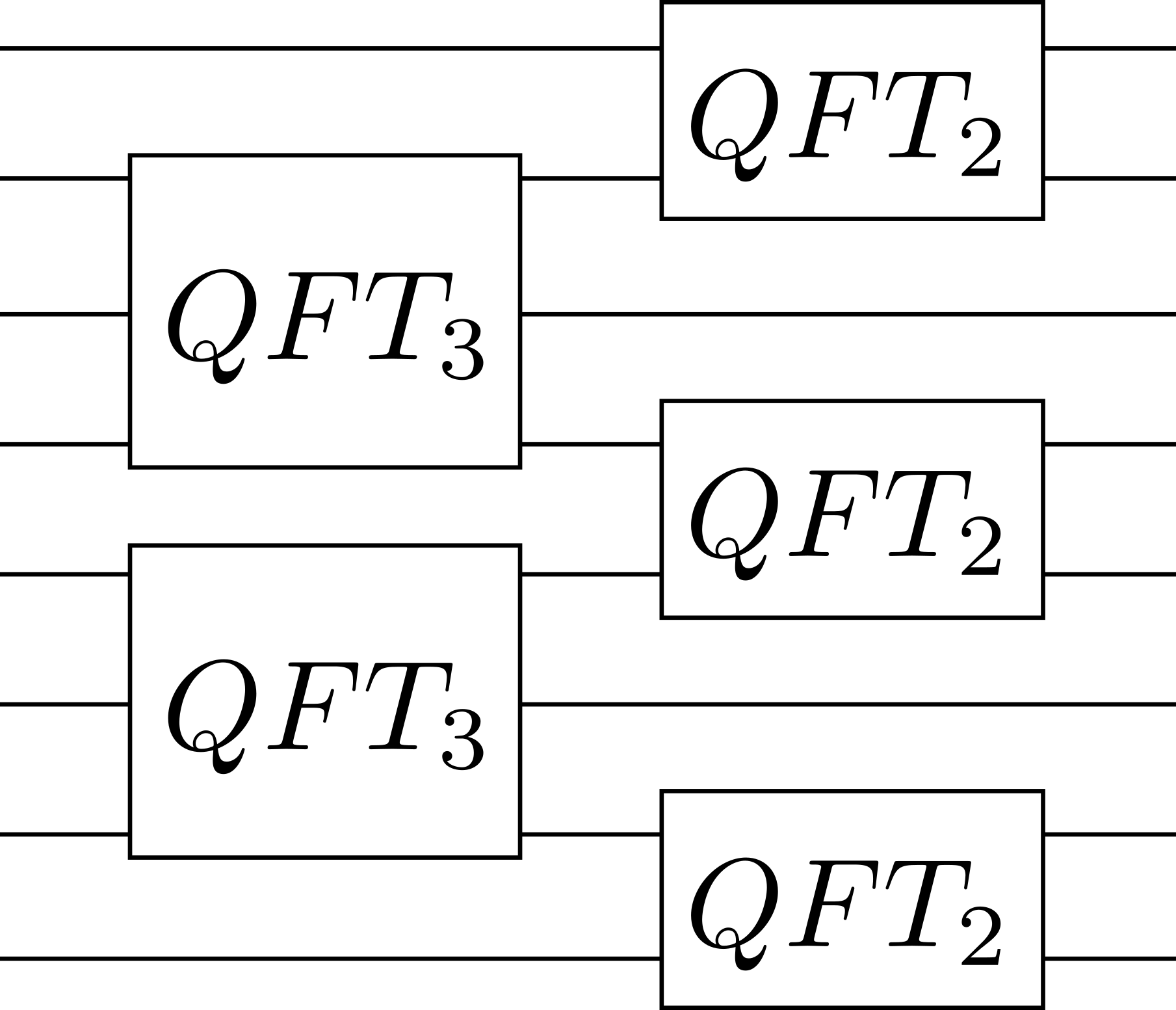}}  & \multirow{2}{*}[-1em]{$\frac{7}{8}=0.8750$}  & $\frac{695}{972} \approx 0.7150$  & $\frac{1}{4}  = 0.2500$ & $\frac{97}{162} \approx 0.5988$   & \multirow{2}{*}[-1em]{$\frac{40319}{40320}\approx 1.0000$}  & $\frac{45095}{46656}  \approx  0.9665$ \\ \cline{2-2}\cline{4-6}\cline{8-8} 
                     & \fbox{\includegraphics[width=0.11\textwidth]{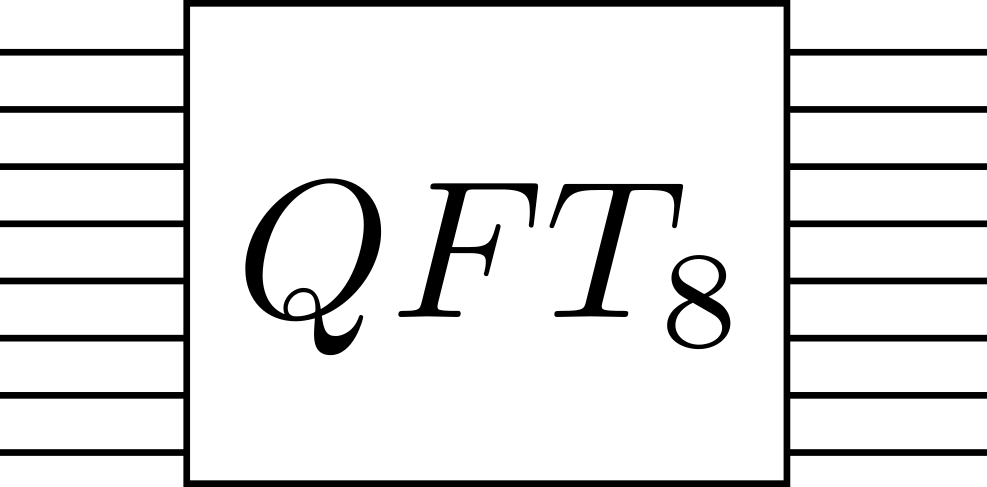}}&                     & \multicolumn{3}{c||}{$\frac{7}{8}=0.8750$} &   & $\frac{7}{8}=0.8750$ \\ \hline
\end{tabular}
\caption{
The best known interferometers for discrimination of the singly and completely distinguishable states of $N = 2$ to $8$ photons in $N$ modes.
For $N=2$ and $3$ the quantum Fourier transform ($QFT_N$) is optimal for both $\rho_\mr{s}$ and $\rho_\mr{d}$, but for $N\geq 4$ the interferometers for each are different; we include all probabilities of success for comparison.
For singly distinguishable states, the quantum Fourier transform saturates the bound and so is optimal for each $N$; due to the $QFT$'s symmetry it does not matter which port the `bad' photon (see Sec.~\ref{subsec:singlyD}) is in, however this is not true of the $\rho_\mr{d}$ interferometers and so we include best, worst and average success probabilities assuming each port is equally likely to be `bad'.
The completely distinguishable state is essentially unique, so there is only one success probability to report; an asterix $\ast$ indicates extensive numerical optimisation leads us to believe the $N=3,4,5$ cases are optimal despite being far from the bound; it is remarkable that the best $\rho_\mr{d}$ interferometers have constant optical depth (made up of $QFT_3$s followed by $QFT_2$s) for each $N$.
Interestingly, the two success probabilities of the $QFT$ are always equal except for $N=6$, the only case in the table that is not a power of a prime.
The measurement outcomes that lead to these probabilities are specified in Table~\ref{tab:clicks}.
}
\label{tab:CDopt}
\end{table*}

\subsubsection{Discriminating singly distinguishable states}
\label{subsec:moreOBE}

Using the numerical optimisation described in Sec.~\ref{subsec:numerical}, we also examined the discrimination of singly distinguishable states for $N=4$ and $5$ photons.
Together with the results for $N=2$ and $3$, we see the optimisation return interferometers equivalent to $QFT_N$, each giving a success probability $1 - 1/N$, saturating the bound in Sec.~\ref{subsec:bounds}.
We have confirmed this behaviour by direct calculation up to $N=9$.

\subsubsection{Discriminating completely distinguishable states}
\label{subsec:fourAndFive}

Numerical optimisation for the $N=4$ and $5$ photon completely distinguishable states yields success probabilities of $19/24$ and $31/36$, respectively.
Both of these are less than the general bounds of Sec.~\ref{subsec:bounds}, ($23/24$ and $35/36$, respectively), and so we cannot conclude they are optimal.
We observe that they do both exceed the singly distinguishable bound of $1-1/N$, consistent with the intuition that it is easier to discriminate a completely distinguishable state than one that is less distinguishable.

The numerics are sensitive to the penalties used in Eq.~(\ref{eq:costfnFD}), due to the existence of interferometers with very similar performance.
For $N=4$, a penalty $\xi=10$ returns an interferometer with success probability $25/32$ that minimises the cost function with a value of $-0.839477$, while a better interferometer with success probability $19/24$ exists but gives a higher value of $-0.836287$.
Increasing the penalty to $50$ yields costs $-0.78455$ and $-0.79277$ for these two interferometers respectively, showing that the latter is now the minimum.
However, increasing the penalty makes optimisation more difficult, because the landscape flattens and gradients go to zero.
For this reason, penalties of $10$, $13$, $15$, $17$, $20$, $25$, $35$ and $50$ were used for $N=4$, and $10$, $12$, $14$, $15$, $16$, $18$, $20$, $35$, and $60$ for $N=5$.

While the complexity of the calculations precluded any further optimisation for $N>5$, we notice that the best interferometers for $N=2,3,4,5$ can be composed out of $QFT_3$ followed by $QFT_2s$.
This suggests a `recursive' structure for the best discriminating interferometers; for $N=6$, $7$, $8$ we tried combinations of $QFT_N$, $QFT_{N-1}$ and so on, and found that discriminators composed of $QFT_3$s followed by $QFT_2$s performed the best.
This is remarkable as these are of constant optical depth, independent of $N$.
Indeed, increasing the optical depth beyond this seems to decrease the success probability, which allowed us to limit our search to a manageable number of configurations.
These are educated guesses however, and do not rule out the existence of better interferometers that might be found.

Table~\ref{tab:CDopt} contains a summary of these results.
We report the probabilities for the best interferometers found to successfully discriminate $\rho_\mr{s}$ and $\rho_\mr{ d}$ from $\rho_\mr{i}$ up to $N=8$.
The measurement outcomes that achieve these probabilities up to $N=5$ are specified in Table~\ref{tab:clicks}, where in the interest of saving space we give the occupations that fail (i.e. correspond to the \emph{ambiguous} POVM element $E_?$) instead of the successful discriminators, because the latter far exceed the former.
For comparison, for each interferometer we include success probabilities for both states of interest to be discriminated from the completely indistinguishable state.
Note that as discussed above for $N=3$, although the $QFT_N$ interferometer is optimal for $\rho_\mr{s}$ no matter which System mode the `bad' photon is in, this will not be true for interferometers that lack the symmetry of $QFT_N$.
Indeed, the best $\rho_\mr{d}$ discriminator does not treat each System mode the same way, and so when using such an interferometer to discriminate $\rho_\mr{s}$ we report best, worst and average success probabilities, assuming each System mode is equally likely to contain the `bad' photon.

\begin{table*}[!t]
\scriptsize
\centering
\begin{tabular}{|c||c|c|}
\hline
$N$ & {\small $\rho_{\mr{s}}$}									& {\small $\rho_{\mr{d}}$} 						\\
\hline
\hline
2 & 20,02											& 20,02								\\
\hline
3 & 300,030,003										& 300,030,003							\\
   & 111											& 111 								\\
\hline
4 & 4000,0400,0040,0004								& 4000,0400,0040,0004						\\
   & 												& 3100,1300,1030,1003,0130,0103				\\
   & 2020,0202										& 									\\
   & 2101,1210,1012,0121								& 2011,0211								\\
\hline
5 & 50000,05000,00500,00050,00005							& 50000,05000,00500,00050,00005				\\
   & 												& 40010,40001,10040,10004,04010,04001,01040,01004	\\
   & 31001,30110,13100,11030,10301,10013,03011,01310,01103,00131	& 31010,31001,13010,13001,10310,10301,10031,10013,01310,01301,01031,01013	\\
   & 22010,21200,20102,20021,12002,10220,02201,02120,01022,00212	& 20120,20102,02120,02102										\\
   & 11111											&														\\
\hline
\end{tabular}
\caption{
Measurement occupations corresponding to the ambiguous POVM element $E_?$ that do \emph{not} discriminate the two states of interest for the numerically optimised interferometers in Table~\ref{tab:CDopt} ($N=2,3,4,5$) -- these are in general far fewer than the number of successful discriminating occupations, and so easier to list.
Recall that for $\rho_{\mr{s}}$, the optimal choice of $QFT_N$ does not depend on the mode in which the single distinguishable photon is present, and neither do the occupations.
Note that although all of the occupations not listed here satisfy Eq.~(\ref{eq:Allcnstrt}), some might have zero probability of occurring and therefore not contribute to discrimination.
}
\label{tab:clicks}
\end{table*}

\section{Discussion and further work}
\label{sec:discuss}

Although we have focused on single and complete distinguishability, as shown in Sec.~\ref{sec:UU} the formalism admits arbitrary states.
Consider for example Fock arrays with a single excitation in each System mode and an arbitrary Label occupation, call it $\underline{n}_\mr{L}$.
Applying the Schur-Weyl transform and focusing on the symmetric irrep $(N)$, where the support is one dimensional, we see that the reduced system state will be of the form
\begin{align}
\frac{\underline{n}_\mr{L}!}{N!} \ket{(N),\underline{1}}\bra{(N),\underline{1}} + \left(1-\frac{\underline{n}_\mr{L}!}{N!}\right) \rho_{\overline{\mr{i}}} .
\label{eq:rhog}
\end{align}
This gives a bound of $1-\underline{n}_\mr{L}!/N!$ on the probability for successfully discriminating such a state from the completely indistinguishable one, and includes the singly and completely distinguishable cases above.
The exact form of such states could be found by reasoning as in Sec.~\ref{sec:states}.

We can use the formalism to compute the number of parameters that describe an arbitrary partially distinguishable collection of $N$ particles in $N$ (or more generally $d$) modes.
Because of the maximal entanglement over $p$ in Eq.~(\ref{eq:lambdasymSL}), when we trace out the Label of an arbitrary totally symmetric state in Eq.~(\ref{eq:symSL}), the resulting mixed state has identical blocks for each copy of $\lambda$, (the number of identical blocks being equal to the outer multiplicity).
Thus the most general mixed state is described by a single (Hermitian) block for each irrep.
Recalling that the number of real parameters in a $d$-dimensional Hermitian matrix is $d^2$, we have for an arbitrary partially distinguishable mixed state of $N$ bosons in $N$ modes (subtracting one for normalisation) $\sum_\lambda {d_{\{\lambda\}}}^2-1={N^2+N-1 \choose N}-1$ real parameters.
If we restrict to coincident input, the number of states is given by the number of \emph{standard} Young tableaux, $d_{(\lambda)}$.
This is because coincidence implies each single particle state is different, and so semistandard tableau become standard; in this case we have $\sum_\lambda {d_{(\lambda)}}^2 - 1 = N!-1$ real parameters.
This number decreases significantly if pure Label states are assumed.
A pure state in $d$ dimensions has $2(d-1)$ real parameters, and every pure state added to a set can add at most one parameter beyond those required to describe its projection onto the $d-1$ dimensional space spanned by the states that came before it.
Thus there are $\sum_{d=2}^N ( 2(d-2)+1) = (N-1)^2$ real parameters in this case, which agrees with previous analyses~\cite{Menssen2017} but is far fewer than the general case.
Note that all of these quantities are of course larger than ${N \choose 2}$, the number of pairwise distinguishabilities classical intuition might lead one to believe are necessary to measure~\cite{Adamson2008}.

There are many other state discrimination scenarios we could consider.
For example, we could try to unambiguously discriminate $\rho_{\mr{d}}$ from $\rho_{\mr{s}}$, or two entirely different states, or more than two states.
Note that due to the `nested' structure of our  three states of interest (cf. Eq.~(\ref{eq:statesetup})), attempting to find a UD POVM $\{E_{\mr{i}}, E_{\mr{d}}, E_{\mr{s}}, E_{?}\}$ reduces to only being able discriminate $\rho_{\mr{d}}$ from the rest.
Another version of discrimination to consider is using bucket (yes/no) instead of number resolving detectors, which are simpler to engineer.
While our focus has been on optimizing over all the possible measurement patterns to obtain the highest possible success probability, as mentioned in Sec.~\ref{subsec:measurements} another type of optimisation that can be carried out is choosing a fixed set of patterns and optimizing the interferometer $U$ only.
The difference would be that in Eqs.~(\ref{eq:Allmax}, \ref{eq:Allcnstrt}) $D$ would now be fixed, simplifying the problem.
As an example, during the preparation of this manuscript a closely related paper was released~\cite{Brod2018}, where the authors study a single reference photon input into a $QFT_{N-1}$, followed by $QFT_2$ HOM tests on the $N-1$ outputs with the rest of the $N-1$ photons (for a total of $N$ photons in $2N-2$ modes).
This is equivalent to a UD procedure where $D$ is fixed as the set of $N$-fold coincidences.
The approach is different and so it is not surprising that it is suboptimal for discrimination, however this interferometer's behaviour is clear for all $N$.

Finally, we have no doubt that proofs for many of the results here, such as $QFT_N$ optimality for discriminating singly distinguishable states, should be possible, but they are left as further work.

The data associated with this paper is available for download at the University of Bristol data repository, data.bris~\cite{data}.

\section*{Acknowledgements}
The authors are pleased to acknowledge helpful conversations with S.~Bartlett, S.~Croke, A.~Doherty, H.~De~Guise, J.~Silverstone, and especially P.~Birchall, S.~Pallister, T.~Rudolph, C.~Sparrow and T.~Sugiyama who undertook early investigations.
SS was supported by the Bristol Quantum Engineering Centre for Doctoral Training, EPSRC grant EP/L015730/1.
PST was supported in part by EPSRC First Grant EP/N014812/1.

\bibliography{discrimination}
\clearpage 

\appendix

\end{document}